\documentclass[twocolumn, twocolappendix]{aastex631}
\usepackage{enumitem}
\usepackage{xspace}
\usepackage{amsmath}
\usepackage{float}

\newcommand{\sw}[1]{\texttt{#1}}

\shorttitle{Connecting the FRB DM--redshift relation to Suppression of the Matter Power Spectrum}
\shortauthors{Sharma et al.}

\begin{document}

\title{A hydrodynamical simulations-based model that connects the FRB DM--redshift relation to suppression of the matter power spectrum via feedback}

\correspondingauthor{Kritti Sharma}
\email{kritti@caltech.edu}

\author[0000-0002-4477-3625]{Kritti Sharma}
\affiliation{Cahill Center for Astronomy and Astrophysics, MC 249-17 California Institute of Technology, Pasadena CA 91125, USA.}

\author{Elisabeth Krause}
\affiliation{Department of Astronomy/Steward Observatory, University of Arizona, 933 North Cherry Avenue, Tucson, AZ 85721, USA}
\affiliation{Department of Physics, University of Arizona, 1118 E Fourth Street, Tucson, AZ 85721,
USA}

\author{Vikram Ravi}
\affiliation{Cahill Center for Astronomy and Astrophysics, MC 249-17 California Institute of Technology, Pasadena CA 91125, USA.}
\affiliation{Owens Valley Radio Observatory, California Institute of Technology, Big Pine CA 93513, USA.}

\author{Robert Reischke}
\affiliation{Argelander-Institut für Astronomie, Universität Bonn, Auf dem Hügel 71, D-53121 Bonn, Germany}

\author{Pranjal R. S.}
\affiliation{Department of Astronomy/Steward Observatory, University of Arizona, 933 North Cherry Avenue, Tucson, AZ 85721, USA}

\author[0000-0002-7587-6352]{Liam Connor}
\affiliation{Center for Astrophysics | Harvard $\&$ Smithsonian, Cambridge, MA 02138-1516, USA.}

\begin{abstract}
Understanding the impact of baryonic feedback on the small-scale ($k \gtrsim 1\,h\,$Mpc$^{-1}$) matter power spectrum is a key astrophysical challenge, and essential for interpreting data from upcoming weak-lensing surveys, which require percent-level accuracy to fully harness their potential. Astrophysical probes, such as the kinematic and thermal Sunyaev-Zel'dovich effects, have been used to constrain feedback at large scales ($k \lesssim 5\,h\,$Mpc$^{-1}$). The sightline-to-sightline variance in the fast radio bursts (FRBs) dispersion measure (DM) correlates with the strength of baryonic feedback and offers unique sensitivity at scales upto $k \sim 10\,h\,$Mpc$^{-1}$. We develop a new simulation-based formalism in which we parameterize the distribution of DM at a given redshift, $p(\mathrm{DM}|z)$, as a log-normal with its first two moments computed analytically in terms of cosmological parameters and the feedback-dependent electron power spectrum $P_\mathrm{ee}(k, z)$. We find that the log-normal parameterization provides an improved description of the $p(\mathrm{DM}|z)$ distribution observed in hydrodynamical simulations as compared to the standard $F$-parameterization. Our model robustly captures the baryonic feedback effects across a wide range of baryonic feedback prescriptions in hydrodynamical simulations, including \texttt{IllustrisTNG}, \texttt{SIMBA} and \texttt{Astrid}. Leveraging simulations incorporates the redshift evolution of the DM variance by construction and facilitates the translation of constrained feedback parameters to the suppression of matter power spectrum relative to gravity-only simulations. We show that with $10^4$ FRBs, the suppression can be constrained to percent-level precision at large scales and $\sim 10$\% precision at scales $k \gtrsim 10\,h\,$Mpc$^{-1}$ with prior-to-posterior $1\sigma$ constraint width ratio $\gtrsim 20$.
\end{abstract}

\section{Introduction} \label{sec:introduction}

Over the past two decades, observational cosmology has established the standard cosmological model, $\Lambda$ cold dark matter ($\Lambda$CDM), in which quantum fluctuations during inflation seed cosmic structure, and a cosmological constant drives late-time accelerated expansion. Observations of the expansion history, obtained through analyses of the fluctuations in the cosmic microwave background~\citep[CMB;][]{2020A&A...641A...6P}, lensing of the CMB at intermediate redshifts~\citep{2020A&A...641A...8P}, Type Ia supernovae as standard candles~\citep{2022ApJ...934L...7R, 2024arXiv240806153F}, the baryon acoustic oscillation (BAO) scale as a standard ruler~\citep{2017MNRAS.470.2617A, 2025JCAP...02..021A}, and time-delay cosmography of gravitationally lensed quasars~\citep{2020MNRAS.498.1420W} and supernovae~\citep{2019ApJ...876..107P}, have constrained the key cosmological parameters, enabling rigorous tests of the standard $\Lambda$CDM model. While the origins of the Hubble tension and the $S_8$ tension\footnote{$S_8= \sigma_8 \Omega_\mathrm{m}^{0.5}$ quantifies the degree of late-time matter clustering, where $\sigma_8$ is the root-mean-square amplitude of the linear matter density fluctuations smoothed with a top-hat filter of comoving radius 8\,$h^{-1}$Mpc and $\Omega_\mathrm{m}$ is the matter density parameter.} -- whether physical or unaccounted systematics -- remain under investigation, the broader consensus questioning the completeness of the standard model continues to build (see \citealt{2025arXiv250401669D} for a review).

Paving the way for the era of precision cosmology, current and upcoming experiments, such as the Vera Rubin Observatory~\citep[][]{2019ApJ...873..111I}, Euclid~\citep{2011arXiv1110.3193L} and Wide-Field Infrared Survey Telescope~\citep[WFIRST;][]{2013arXiv1305.5425S, 2018arXiv181200514F} are designed to map the distribution of matter with unprecedented precision, leveraging techniques like gravitational lensing\footnote{The galaxy clustering analyses are primarily limited by the modeling of the non-linear galaxy bias (as opposed to baryonic feedback). Currently, the analyses using perturbation theory-based models are limited to $k \sim 0.25\,h\,$Mpc$^{-1}$, where the impact of feedback is almost negligible.}~\citep{2022PhRvD.105b3514A, 2022PhRvD.105b3515S, 2023OJAp....6E..36D, 2023PhRvD.108l3519D, 2023PhRvD.108l3517M, 2023A&A...675A.189D, 2025arXiv250319441W}.

To meet the expected statistical precision of these experiments, the impact of baryonic feedback on the matter power spectrum needs to be constrained with percent-level precision at scales $k \lesssim 10\,h$\,Mpc$^{-1}$~\citep{2012JCAP...04..034H, 2022PhRvD.105b3514A, 2023MNRAS.525.5554P, 2024JCAP...07..037T}. The cosmic shear measurements at such small scales should practically only be limited by shape noise. Ignoring the impact of baryons can introduce systematic uncertainties into cosmological analyses, where the expected bias in the parameters of the equation of state ($w_0, w_a$) and the sum of neutrino masses can be as large as $\sim 5\sigma$ and $\sim 20\sigma$, respectively~\citep{2019OJAp....2E...4C}. Therefore, cosmic shear analyses would benefit from accurate models of the matter power spectrum at small scales for cosmological inference.

Broadly, cosmic shear analyses so far internally mitigate baryonic feedback through (i) scale cuts excluding scales affected by feedback~\citep{2018PhRvD..98d3528T, 2024arXiv241112022D}, (ii) analytic halo models with empirical baryon profiles~\citep{2020A&A...641A.130M, 2024arXiv240118072P} (iii) power spectrum emulators trained on post-processed N-body simulations~\citep[the “baryonification” approach;][]{2019JCAP...03..020S, 2021MNRAS.506.4070A, 2023MNRAS.520.3626W}, and (iv) data-driven marginalization, such as principal component analysis~\citep[PCA;][]{2015MNRAS.454.2451E}. These approaches eliminate systematics by either sacrificing valuable information from non-linear scales or marginalizing over complex feedback models, thus reducing the constraining power.

Complementary to these strategies, independent observational probes of baryons offer a promising avenue for constraining the effects of baryonic feedback. For example, joint analyses of cosmic shear with X-ray observations~\citep[e.g.][]{2022MNRAS.514.3802S, 2024PhRvL.133e1001F, 2024A&A...689A.298G, 2024MNRAS.528.4379G, 2025ApJ...983....8L}, the kinematic Sunyaev-Zel’dovich~\citep[kSZ; e.g.][]{2021PhRvD.104d3502C, 2022MNRAS.514.3802S, 2024MNRAS.534..655B, 2024arXiv240707152H, 2025arXiv250116983B} effect and the thermal Sunyaev-Zel’dovich~\citep[tSZ; e.g.][]{2022A&A...660A..27T, 2023MNRAS.525.1779P, 2024OJAp....7E..13B} effect help constrain the impact of baryonic feedback at small scales, using the information from independent probes of the baryon distribution. 

Fast Radio Bursts~\citep[FRBs;][]{2022A&ARv..30....2P} are a class of millisecond-duration transients that probe the distribution of baryons at scales upto $k \sim 10\,h\,$Mpc$^{-1}$ (see \S\ref{subsec:relevant_scales} for a description). The frequency-dependent dispersive delays caused by intervening plasma are encoded in the spectrum of the burst, quantified as the dispersion measure (DM), which probes the ionized electron column density along the line of sight, thus helping us constrain the location of baryons~\citep{2020Natur.581..391M}. Specifically, the sightline-to-sightline variance in FRB DMs, parameterized by the form of the DM distribution $p(\mathrm{DM}|z)$, is sensitive to the influence of feedback on the degree of clustering of baryons~\citep[e.g.,][]{2014ApJ...780L..33M, 2022ApJ...928....9L, 2023arXiv230909766R, 2024arXiv240200505K, 2024arXiv240916952C, 2024arXiv241117682R, 2025arXiv250117922M}.

FRBs were quickly recognized as powerful probes of intergalactic baryons, with early studies highlighting their potential for cosmological applications~\citep{2007Sci...318..777L, 2013Sci...341...53T, 2014ApJ...780L..33M}. The first precise localizations~\citep{2017Natur.541...58C, 2019Natur.572..352R, 2019Sci...365..565B} established the extragalactic distance scale, and \citet{2020Natur.581..391M} was the first to model the DM-redshift distribution of an FRB sample, laying the foundation for using FRBs as large-scale baryon tracers. Combining FRBs with other cosmological probes, such as Type Ia supernovae, BAO, and CMB, \citet{2023ApJ...944...50W} measured the fraction of diffuse baryons in the Universe. More recently, \citet{2024arXiv240916952C} and \citet{2024arXiv240200505K} advanced this approach by accurately partitioning baryons between halos and the cosmic web, refining our understanding of the cosmic matter distribution.

With the advent of next-generation instruments such as the Deep Synoptic Array~\citep[DSA-2000;][]{2019BAAS...51g.255H} and the Canadian Hydrogen Observatory and Radio-Transient Detector~\citep[CHORD;][]{2019clrp.2020...28V}, the field of FRB science is transitioning into an era wherein tens of thousands of FRBs will be detected and localized by these experiments. This necessitates a reassessment of existing methodologies for modeling and interpreting the observed $p(\mathrm{DM}|z)$ distribution to fully harness the potential of FRBs as cosmological probes. 

\begin{figure}
\centering
\includegraphics[width=\columnwidth]{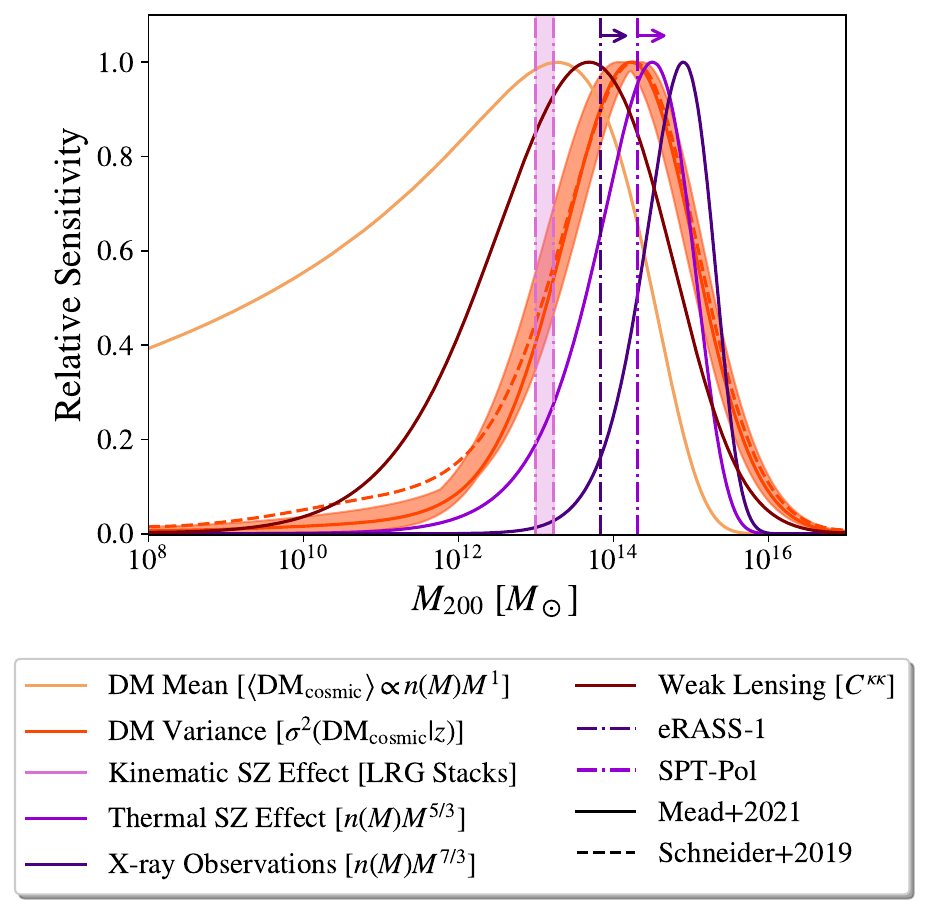}
\caption{Halo mass sensitivity of various baryons tracers, including X-ray observations, thermal Sunyaev-Zel'dovich (tSZ) effect, kinematic Sunyaev-Zel'dovich (kSZ) effect, cosmic shear and FRB dispersion measure (DM). While the mean DM is sensitive to $\gtrsim 10^8\,M_\odot$ halos (neglecting feedback effects that may evacuate small halos), the DM variance depends on the degree of clustering of baryons in halos and is sensitive to more massive halos ($\gtrsim 10^{11}\,M_\odot$) with peak sensitivity at $\sim 10^{14}\,M_\odot$. This sensitivity of DM variance is evaluated using the halo model of \citet{2021MNRAS.502.1401M} under two scenarios: (i) ejected gas traces dark matter at large scales and (ii) ejected gas follows profiles of \citet{2019JCAP...03..020S}, as implemented in \sw{BaryonForge}~\citep{2024OJAp....7E.108A}. The relative contribution of halos depends on model parameters governing the fraction of bound gas and halo concentration, the variations of which may span the shaded region.}
\label{fig:sensitivity}
\end{figure}

We illustrate the complimentary nature of FRBs as an additional probe by evaluating the halo mass sensitivity of FRB observables and comparing them with other tracers of baryons in Figure~\ref{fig:sensitivity} (see Appendix~\ref{appendix:halo_sensitivity} for a description of the sensitivity evaluation methods). The DM is sensitive to all baryons along the line of sight and under the standard halo model assumption that all mass is partitioned into halos\footnote{Practically, most of the DM comes from non-virialized objects~\citep{2024A&A...683A..71W}.}, its sensitivity scales as $~\sim n(M) M$, where $n(M)$ is the halo mass function. This is a weaker dependence on mass compared to other probes\footnote{Technically, the sensitivity scales as $\sim M f_\mathrm{gas}(M) n(M)$, where $f_\mathrm{gas}(M)$ represents the fraction of gas retained by halos, depending on the strength of feedback. For simplicity, we omit $f_\mathrm{gas}(M)$ for our pilot study.}. The key FRB observable that correlates with the suppression of the matter power spectrum is the variance in the DM across FRB events at the same redshift but different sightlines. While the mean DM simply scales with halo mass, the variance depends on the clustering statistics of baryons (see \S\,\ref{sec:FRB_DM_formalism} and Appendix~\ref{appendix:halo_sensitivity}). We find that the halo mass sensitivity of the  variance peaks in massive $\gtrsim 10^{14}\,M_\odot$ halos, with sensitivity down to $\sim 10^{11}\,M_\odot$ halos.

Traditionally, the influence of baryonic feedback in FRB DM analyses has been parameterized using the feedback $F$-parameter~\citep[][]{2020Natur.581..391M}. However, this parameter lacks a rigorous theoretical foundation and has limited effectiveness in fitting the DM distribution in hydrodynamical simulations (see \S\ref{sec:DMz_parameterization}). Other methods include quantifying baryonic effects using the fraction of baryons in the intergalactic medium (IGM) and the circumgalactic medium (CGM) of intervening halos, where \citet{2024arXiv240200505K} map the foreground of FRB sightlines using baryon density reconstruction (smoothed by a Gaussian kernel, the size of which is calibrated to Illustris~\citep{2014MNRAS.444.1518V}), whereas \citet{2024arXiv240916952C} jointly model the DM distribution in the IGM and CGM as a bi-variate log-normal distribution (calibrated to \texttt{IllustrisTNG}~\citep{2019ComAC...6....2N}).

In this study, we develop a model mapping the variance of $p(\mathrm{DM}|z)$ distribution to the suppression of matter power spectrum, which is calibrated to the \texttt{IllustrisTNG} suite of 1,000 hydrodynamical simulations in the CAMELS project~\citep{2023ApJS..265...54V, 2023ApJ...959..136N}, spanning a broad range of baryonic feedback scenarios. Specifically, we construct the $p(\mathrm{DM}|z)$ distribution with a log-normal parameterization, which is fully defined by its mean (see Equation~\ref{eqn:meanDM}) and variance (see Equation~\ref{eqn:variance}), where the latter is computed in terms of cosmology and feedback-dependent electron power spectrum, $P_\mathrm{ee}(k,z)$. Given a statistical sample of FRBs, we construct a likelihood for the distribution of DM values as a function of redshift, and perform a joint inference of both cosmological and feedback parameters using Markov Chain Monte Carlo (MCMC) methods. 

Modeling DM variance within a given feedback model in hydrodynamical simulations allows an accurate representation of the redshift evolution of DM variance, while simultaneously linking constrained feedback parameters to the suppression of the matter power spectrum. We demonstrate that, using our formalism, an idealized sample of $10^4$ FRBs can constrain the impact of baryons on the matter power spectrum with percent-level precision at large scales and with $\sim$10\% precision at small scales, thus providing crucial inputs for future weak lensing analyses. These constraints can be used to infer the baryon fractions of galaxy groups and clusters~\citep{2024arXiv240905716S}. In parallel, these measurements also provide a basis for interpreting gas profiles within the standard halo model framework~\citep{2020A&A...641A.130M, 2021MNRAS.502.1401M}.

This article is structured as follows. We review the theoretical foundations of our formalism in \S\ref{sec:FRB_DM_formalism}. We compare the effectiveness of the \citet{2020Natur.581..391M} parameterization with the log-normal parameterization of the $p(\mathrm{DM}|z)$ distribution in \S\ref{sec:DMz_parameterization}. We formulate the $P_\mathrm{ee}(k,z)$, and consequently the DM variance, as a function of feedback parameters using the \texttt{IllustrisTNG} suite of hydrodynamical simulations in \S\ref{sec:feedback_sims}. In \S\ref{sec:simulation_based_inference}, we discuss our \texttt{IllustrisTNG}-calibrated, hydrodynamical simulation-based model of the DM$-z$ relation, assess its adaptability to varied feedback implementations across different simulations, such as \texttt{SIMBA}~\citep{2019MNRAS.486.2827D} and \texttt{Astrid}~\citep{2022MNRAS.512.3703B}, and discuss the limitations of our methodology. Finally, we summarize our findings in \S\ref{sec:conclusions}.

\section{FRB Dispersion Measure Formalism} \label{sec:FRB_DM_formalism}

The propagation of a radio pulse with frequency $\nu$ through the cosmic plasma of free electrons results in a delay in its arrival time, which is proportional to $\nu^{-2}$. This frequency-dependent delay, known as the dispersion delay, is characterized by the DM. For an FRB source at redshift $z_\mathrm{s}$ in the direction $\hat{x}$, the $\mathrm{DM}_\mathrm{FRB}$ is given by the integral of the electron number density ($n_\mathrm{e}$) along the line of sight, weighted by $(1+z)^{-1}$, 
\begin{equation}
\mathrm{DM}_\mathrm{FRB} (z_\mathrm{s},~\hat{x}) = \int\limits_0^{\mathrm{s}(z_\mathrm{s})} \frac{n_\mathrm{e} (z,~\hat{x})}{(1+z)} \mathrm{d}s,
\label{eqn:DM_FRB_definition}
\end{equation}
where $\mathrm{d}s = \mathrm{d}\chi/(1+z)$ is the differential of the proper distance, $\mathrm{d}\chi = c~\mathrm{d}z/H(z)$ is the differential of the comoving distance, and $H(z)$ is the Hubble parameter. Progressing along the line of sight from the observer to the FRB source, the $\mathrm{DM}_\mathrm{FRB}$ can be written as: 
\begin{equation}
\begin{split}
    \mathrm{DM}_\mathrm{FRB} (z_\mathrm{s},~\hat{x})  = &\;\mathrm{DM}_\mathrm{MW}(\hat{x}) + \mathrm{DM}_\mathrm{cosmic}(z_\mathrm{s},~\hat{x})  \\& +  \frac{\mathrm{DM}_\mathrm{host}}{1+z_\mathrm{s}}.
    \label{eqn:DM_FRB_split}
\end{split}
\end{equation}
Here, the Milky Way component ($\mathrm{DM}_\mathrm{MW}$), includes the contribution of the interstellar medium, which is often computed using the NE2001 model~\citep{2002astro.ph..7156C, 2003astro.ph..1598C}, and the galactic halo~\citep{2019MNRAS.485..648P, 2023arXiv230101000R}.

The last two terms in Equation~\ref{eqn:DM_FRB_split} represent the extragalactic DM contribution ($\mathrm{DM}_\mathrm{exgal}$). The host galaxy component (DM$_\mathrm{host}$) includes the contribution of the circumburst medium, interstellar medium, and halo of the host galaxy. Simulations have shown that the DM$_\mathrm{host}$ distribution is well approximated as log-normal~\citep{2020ApJ...900..170Z, 2024arXiv240308611T}: 
\begin{equation}
    \begin{aligned}
        & p(\mathrm{DM}_\mathrm{host}) \\
        & = \frac{1}{\mathrm{DM}_\mathrm{host}\sigma_\mathrm{host} \sqrt{2\pi}}\mathrm{exp}\left( - \frac{(\ln \mathrm{DM}_\mathrm{host} - \mu_\mathrm{host})^2}{2\sigma_\mathrm{host}^2} \right),
    \end{aligned}
    \label{eqn:DMhost}
\end{equation}
with parameters $(\mu_\mathrm{host}, \sigma_\mathrm{host})$, such that $\langle \mathrm{DM}_\mathrm{host} \rangle = e^{\mu_\mathrm{host}}$ and $\sigma^2[\mathrm{DM}_\mathrm{host}] = e^{2\mu_\mathrm{host} + \sigma_\mathrm{host}^2}(e^{\sigma_\mathrm{host}^2}-1)$ are the mean and variance of the distribution, respectively. The zoom-in simulations of galaxies suggest that high DM$_\mathrm{host}$ may be expected if FRBs trace star-formation~\citep{2024ApJ...972L..26O} and these are consistent with the observed DM$_\mathrm{host}$ of FRBs, which is well approximated by a log-normal distribution with $\mu_\mathrm{host} \sim 5$ and $\sigma_\mathrm{host} \sim 0.5$~\citep{2024arXiv240916952C}.

The cosmic component ($\mathrm{DM}_\mathrm{cosmic}$), includes the contribution of the IGM and the CGM of intervening halos. The closed-form expression for the Fourier transform of $p(\mathrm{DM}_\mathrm{cosmic}|z)$ depends on the halo gas profile~\citep{2007ApJ...671...14Z, 2014ApJ...780L..33M} and constraining the gas density profiles by stacking FRB sightlines for a range of halo impact parameters could be a promising avenue~\citep{2014ApJ...780L..33M}. Currently, very little is known about the gas density profiles of $\sim$ Milky Way-size halos, so an analytical estimate of the low-order moments of $p(\mathrm{DM}_\mathrm{cosmic}|z)$, such as the mean and variance, can guide us towards understanding this distribution. Although higher moments are also required to fully specify $p(\mathrm{DM}_\mathrm{cosmic}|z)$, in \S\ref{sec:DMz_parameterization} we show that a log-normal distribution provides an excellent approximation for the distribution of $\mathrm{DM}_\mathrm{cosmic}$ in hydrodynamical simulations.

\subsection{Mean Dispersion Measure} \label{subsec:meanDM}

For a flat Universe with matter and dark energy, the redshift evolution of the Hubble parameter is given by $H(z) = H_0 \sqrt{\Omega_\mathrm{m}(1+z)^3 + \Omega_\Lambda}$, where $H_0$ is the Hubble constant, $\Omega_\mathrm{m}$ and $\Omega_\Lambda$ are the present epoch matter and dark energy densities, respectively. The mean electron density,
\begin{equation}
    \bar{n}_\mathrm{e}(z) = \frac{f_\mathrm{d}(z) \rho_\mathrm{b}(z)\chi_\mathrm{e}}{m_\mathrm{p}},
\end{equation} 
where $f_\mathrm{d}(z)$ is the fraction of baryons in the diffuse ionized gas, $\chi_\mathrm{e} = Y_\mathrm{H} + Y_\mathrm{He}/2 \approx 1-Y_\mathrm{He}/2$ is the electron fraction computed using the primordial abundance of helium which is constrained to $\sim 1$\% precision by the CMB experiments~\citep{2020A&A...641A...6P} and $m_\mathrm{p}$ is the proton mass. The baryon mass density is $\rho_\mathrm{b}(z) = \Omega_\mathrm{b} \rho_\mathrm{c}(1+z)^3$, where $\Omega_\mathrm{b}$ is the cosmic matter density in baryons and $\rho_\mathrm{c} = 3H_0^2/8\pi G$ is the critical density of the Universe. Substituting these into Equation~\ref{eqn:DM_FRB_definition}, the mean $\mathrm{DM}_\mathrm{cosmic}$ is given by
\begin{equation}
\begin{aligned}
    \langle \mathrm{DM}_\mathrm{cosmic} (z_\mathrm{s}) \rangle & = \int\limits_0^{z_\mathrm{s}} \frac{3c \chi_\mathrm{e} \Omega_\mathrm{b} H_0}{8 \pi G m_\mathrm{p}} \frac{f_\mathrm{d}(z) (1+z) \mathrm{d}z}{\sqrt{\Omega_\mathrm{m}(1+z)^3 + \Omega_\Lambda}} \\
    & = \int\limits_0^{\chi(z_s)} W_\mathrm{DM}(\chi) \mathrm{d}\chi,
\label{eqn:meanDM}
\end{aligned}
    \vspace{-0.1cm}
\end{equation}
which has an explicit dependence on cosmological parameters. The second equality is the equivalent integral using the comoving distance ($\chi$), where the prefactors have been folded into the radial weight function for projected electron density field, $W_\mathrm{DM}(\chi)$.

\subsection{Sightline-to-Sightline Variance} \label{subsec:DMvariance}

The scatter in the $\mathrm{DM}_\mathrm{cosmic}$ at a certain redshift $z$ arises from the collapsed systems encountered along the sightline. The electron number density at redshift $z$ in the direction $\hat{x}$ can be written as the mean density plus a fluctuation term, 
\begin{equation}
    n_\mathrm{e} (z, \hat{x}) = \bar{n}_\mathrm{e} (z) (1+\delta_\mathrm{e}(z, \hat{x})) = \bar{n}_{\mathrm{e},0} (1+z)^3 (1+\delta_\mathrm{e}(z, \hat{x})),
\end{equation}
where $\delta_\mathrm{e}$ is the dimensionless electron density contrast. Substituting this into Equation~\ref{eqn:DM_FRB_definition}, the variance in $\mathrm{DM}_\mathrm{cosmic}$ induced by the large scale structure for an FRB located at redshift $z_\mathrm{s}$ is given by (see \citet{2023MNRAS.524.2237R} for detailed derivation)
\begin{equation}
    \begin{aligned}
    \sigma^2 [\mathrm{DM}_\mathrm{cosmic}(z_\mathrm{s})] & = \sum_\ell \frac{2\ell + 1}{4\pi} C_\mathrm{DM}(\ell, z_\mathrm{s}) \\
    & \approx \int \frac{\ell \mathrm{d}\ell}{2\pi} C_\mathrm{DM}(\ell, z_\mathrm{s}), 
    \end{aligned}
    \label{eqn:covariance}
\end{equation}
where the second equality follows from the flat sky approximation, $C_\mathrm{DM}(\ell, z_\mathrm{s})$ is the angular power spectrum,
\begin{equation}
\begin{aligned}
    C_\mathrm{DM}(&\ell, z_\mathrm{s}) \\
    = &\int\limits_0^\infty k^2 \mathrm{d}k \left( \int\limits_0^{\chi_\mathrm{s}} \mathrm{d}\chi W_\mathrm{DM}(\chi) \sqrt{P_\mathrm{ee} (k, z(\chi))} j_\ell(k\chi)\right)^2,
\end{aligned} 
\end{equation}
$P_\mathrm{ee}(k,z)$ is the electron power spectrum defined as
\begin{equation}
    \langle \tilde{\delta}_\mathrm{e}(k, z) \tilde{\delta}_\mathrm{e}(k^\prime, z) \rangle = (2\pi)^3 \delta_\mathrm{D}(k+k^\prime) P_\mathrm{ee}(k,z),
\end{equation}
$\tilde{\delta}_\mathrm{e}$ is the Fourier transform of the density field, $\delta_\mathrm{D}$ is the Dirac delta function and $j_\ell$ denotes the spherical Bessel function of zeroth order. For sufficiently large $\ell$, the spherical Bessel functions are highly peaked around $k\chi = \ell$ and therefore, $j_\ell(k\chi) \approx \delta_\mathrm{D}(k-\ell/\chi)/k\chi$. Using this Limber approximation,
\begin{equation}
    C_\mathrm{DM}(\ell, z_\mathrm{s}) \approx \int\limits_0^{\chi_\mathrm{s}} \mathrm{d}\chi \frac{W_\mathrm{DM}^2(\chi)}{\chi^2} P_\mathrm{ee}\left(k = \frac{\ell+1/2}{\chi}, z(\chi)\right).
    \label{eqn:limber_approximation}
\end{equation}
Substituting Equation~\ref{eqn:limber_approximation} into Equation~\ref{eqn:covariance}, the variance in $\mathrm{DM}_\mathrm{cosmic}$~\citep{2014ApJ...780L..33M} is given by
\begin{equation}
\begin{aligned}
    \sigma^2 [\mathrm{DM}_\mathrm{cosmic}(z_\mathrm{s})] & = \int\limits_0^{\chi_\mathrm{s}} \mathrm{d}\chi W_\mathrm{DM}^2 \int\limits_0^\infty \frac{k \mathrm{d}k}{2\pi} P_\mathrm{ee} (k, z(\chi)).
\end{aligned}
\label{eqn:variance}
\end{equation}
Although we do not account for covariance between sightlines due to the large scale structure in this work, we note that accounting for it will be crucial to carry out an unbiased inference using large samples of observed FRBs~\citep{2023MNRAS.524.2237R}.

\section{p(DM$|$z) Parameterization} \label{sec:DMz_parameterization}

In the preceding section, we established the analytical framework and outlined the parameterization of the mean and variance of the FRB DM distribution, corresponding to the first and second moments of the probability distribution function $p(\mathrm{DM}_\mathrm{cosmic}|z)$. For likelihood analysis, a first-order approximation often assumes a Gaussian likelihood~\citep{2022MNRAS.511..662H}; however, this assumption is inadequate due to the well-documented non-Gaussian nature of $p(\mathrm{DM}_\mathrm{cosmic}|z)$~\citep{2024arXiv241007084K}. In this section, we examine the \citet{2020Natur.581..391M} parameterization of $p(\mathrm{DM}_\mathrm{cosmic}|z)$ (see \S\ref{subsec:macquart_parameterization}) using FRB sightlines derived from hydrodynamical simulations. Next, we introduce our proposed log-normal parameterization and demonstrate its enhanced performance relative to the \citet{2020Natur.581..391M} parameterization (see \S\ref{subsec:lognormal_parameterization}).

\subsection{Macquart Parameterization} \label{subsec:macquart_parameterization}

Motivated by theoretical studies of the IGM~\citep{2000ApJ...530....1M}, which provides a fitting formula to the results of numerical IGM simulations, \citet{2020Natur.581..391M} parameterized the $p(\mathrm{DM}_\mathrm{cosmic}|z)$ distribution such that it captures the large skewness arising from the intersection of a few structures along the line of sight, which results in an increase in the variance of DM$_\mathrm{cosmic}$:
\begin{equation}
    p(\Delta | z) = A \Delta^{-\beta} \exp \left( - \frac{\Delta^{-\alpha} - C_0}{2 \alpha^2 \sigma_\mathrm{DM}^2} \right),
    \label{eqn:macquart_expression}
\end{equation}
where $\Delta = \mathrm{DM}_\mathrm{cosmic}/\langle \mathrm{DM}_\mathrm{cosmic} (z) \rangle$, $\alpha=3$ and $\beta=3$ are chosen to match the observed inner halo density profiles, $C_0$ is a constant that ensures $\langle \Delta \rangle = 1$ and $\sigma_\mathrm{DM}$ is parameterized as $\sigma_\mathrm{DM} = F z^{-1/2}$. This particular form of $\sigma_\mathrm{DM}$ was originally motivated by the Poisson nature of the number of intersecting halos (each contributing $\mathrm{DM}_\mathrm{halo}$), thus implying $\langle \mathrm{DM}_\mathrm{cosmic} \rangle \approx \bar{n}_e c z/H_0$ and $\sigma_\mathrm{DM} \approx \mathrm{DM}_\mathrm{halo} \sqrt{N}/\langle \mathrm{DM}_\mathrm{cosmic} \rangle \propto z^{-1/2}$. The proportionality constant $F$ is the ``feedback'' parameter. We note that technically, $\sigma_\mathrm{DM}$ is \textit{not} the formal variance of the probability distribution function in Equation~\ref{eqn:macquart_expression} and writing the closed-form expression for its variance is non-trivial. However, the $\sigma_\mathrm{DM}$ potentially correlates with the formal variance of this expression, where in the weak (strong) feedback scenario, the baryons in the halo gas are more concentrated (diffuse), implying a larger (smaller) scatter in the DM$-z$ relation, and hence a larger (smaller) $F$-parameter.

\subsubsection{Regarding the definition of the F-parameter}

We identify several conceptual inconsistencies with the function defined above. Firstly, the assumption of Poisson statistics for the number of intervening structures is no longer valid in the regime where filamentary structures on scales of hundreds of kiloparsecs significantly contribute to the variance in $\mathrm{DM}_\mathrm{cosmic}$~\citep[see \S\ref{subsec:relevant_scales};][]{2024A&A...683A..71W}. Secondly, the parameterization $\sigma_\mathrm{DM} = F z^{-1/2}$ becomes invalid in the non-Euclidean limit of space-time (i.e., at high $z$, the curvature of the Universe becomes visible), thereby necessitating a more accurate parameterization. Furthermore, the ability of $F$-parameter to capture the redshift evolution of the variance in DM$_\mathrm{cosmic}$ is also uncertain~\citep{2024ApJ...965...57B}. Finally, in contrast to the findings of \citet{2021ApJ...906...49Z}, we find that this functional form provides a poor fit to FRB sightlines sampled from hydrodynamical simulations, as we discuss below in detail.

\begin{figure}
\centering
\includegraphics[width=\columnwidth]{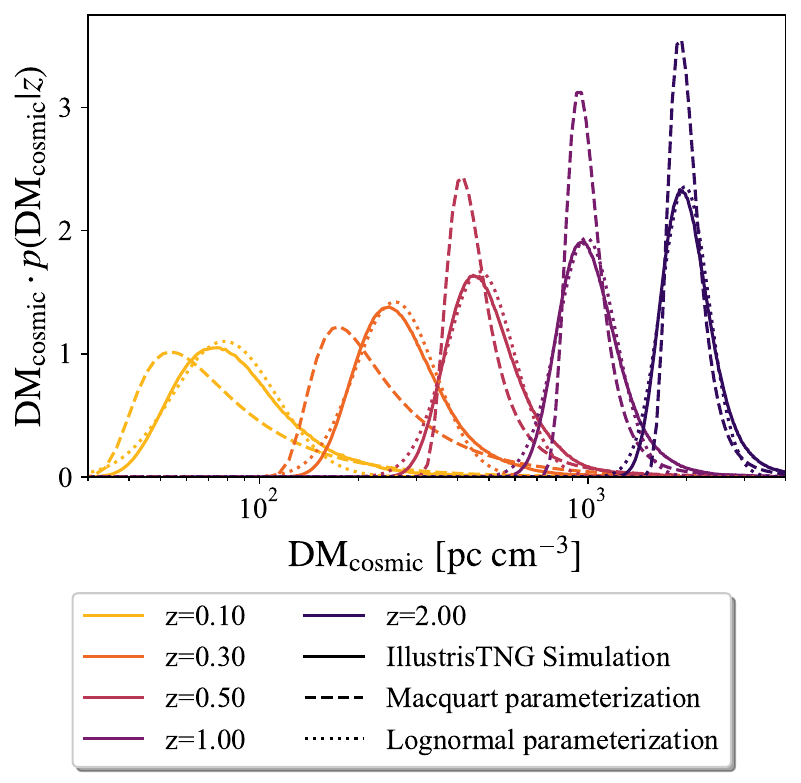}
\caption{Evaluating the efficacy of the $F$-parameter $p(\mathrm{DM}_\mathrm{cosmic} | z)$ parameterization introduced by \citet{2020Natur.581..391M} in comparison to the log-normal parameterization proposed in this work. The comparison is done using the $p(\mathrm{DM}_\mathrm{cosmic} | z)$ distributions derived from the \texttt{IllustrisTNG} simulation within a 205\,$h^{-1}$cMpc box (solid lines), as computed by \citet{2021ApJ...906...49Z}. The log-normal distribution (dotted lines) provides a better fit to the simulated data relative to the $F$-parameter $p(\mathrm{DM}_\mathrm{cosmic} | z)$ parameterization (dashed curves).}
\label{fig:macquart_lognormal_distribution}
\end{figure}

By the definitions of $\Delta$ and $p(\Delta | z)$ in Equation~\ref{eqn:macquart_expression}, the first moment of the distribution (i.e., the mean) must be equal to one. This is mathematically expressed as
\begin{equation}
\frac{\int\limits_0^\infty \Delta p(\Delta) \, \mathrm{d}\Delta}{\int\limits_0^\infty p(\Delta) \, \mathrm{d}\Delta} = 1,\label{eqn:fix_normalization}
\end{equation}
which ensures proper normalization. However, we find that the fits presented in Table~1 and Figure~2 of \citet{2021ApJ...906...49Z} do not satisfy this normalization (also discussed with Zijian Zhang and Daohong Gao via private communication). Consequently, we reanalyze their data and provide the updated fits, represented by the dashed lines in Figure~\ref{fig:macquart_lognormal_distribution}. Upon closer inspection, we find that this functional form poorly fits the expected DM distribution derived from hydrodynamical simulations (shown by solid lines), thus highlighting the need for a simpler functional form that not only provides a better fit but also offers clearer physical interpretability in terms of measurable quantities, as we elaborate in next section.

\begin{figure*}
\centering
\includegraphics[width=\columnwidth]{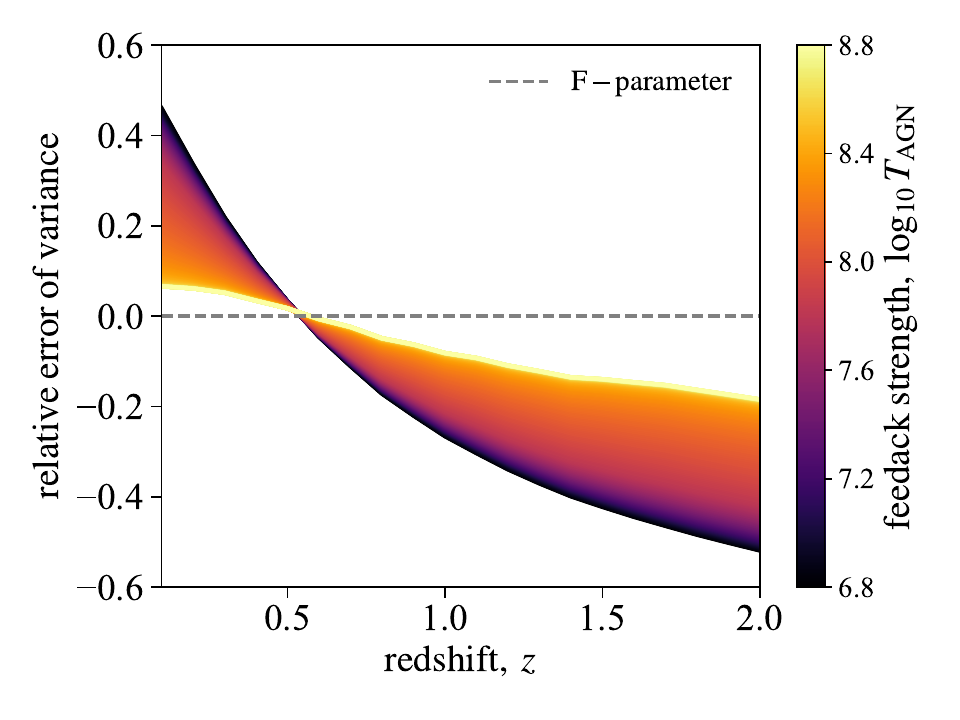}
\includegraphics[width=\columnwidth]{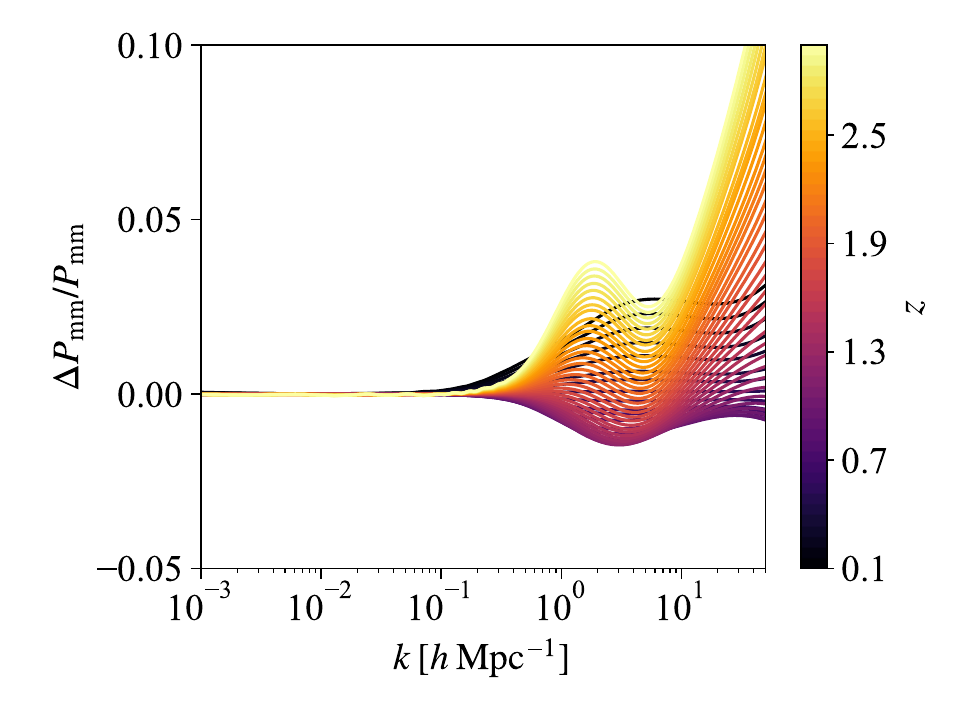}
\caption{The influence of the $F$-parametrization on the estimation of feedback. On the left, we show the relative error introduced by the $F$-parametrization on the sightline-to-sightline variance of the cosmological dispersion measure. If the $F$- parametrization was accurate, all lines should coincide with zero. The right side propagates this error into the matter power spectrum using Equation~\ref{eq:response}. The errors introduced by the $F$-parameterization are significantly larger than the target precision at scales $k \gtrsim 1\,h$\,Mpc$^{-1}$ for upcoming weak lensing surveys, thus necessitating a better parameterization.}
\label{fig:F_parameter_problems}
\end{figure*}

\subsubsection{Regarding the interpretation of the F-parameter}

Here, we quantitatively elaborate on our concerns with the accuracy of physically interpreting the $F$-parameter. To this end, we calculate the variance as in Equation~\ref{eqn:variance} using the publicly available code\footnote{\href{https://github.com/rreischke/frb_covariance}{https://github.com/rreischke/frb\_covariance}} from \citet{2023MNRAS.524.2237R} which uses the feedback model implemented in \texttt{HMCode} \citep{2015MNRAS.454.1958M, 2021MNRAS.502.1401M} emulated with \texttt{cosmopower} \citep{2022MNRAS.511.1771S}. This model emulates the matter power spectrum and the power spectrum of the free electrons controlled by a single parameter, $\log_{10} T_\mathrm{AGN}$, calibrated to the \texttt{BAHAMAS} simulations \citep{2017MNRAS.465.2936M, 2018MNRAS.476.2999M}\footnote{The active galactic nucleus (AGN) heating temperature ($T_\mathrm{AGN}$) correlates with the feedback-driven modifications to halo profile concentration parameters for gas-poor and gas-rich halos, halo-bound gas fractions, and the equation of state of the gas \citep{2020A&A...641A.130M}. Formally, it is related to the \texttt{BAHAMAS} subgrid heating parameter $\Delta T_\mathrm{heat}$ which controls the occurrence of AGN feedback such that it activates only after assembling sufficient energy to heat a fixed number of gas particles by $\Delta T_\mathrm{heat}$}.

We obtain an estimate for the relative variance $\sigma^2(\Delta) = \sigma^2 (z)/\mathrm{DM}^2_\mathrm{cosmic}(z)$ and fit the $F$-parameter relation to it, inversely weighted by an FRB source distribution $n(z)\propto z^2\exp(-\alpha z)$, with $\alpha = 3.5$ to mimic the availability of FRBs as a function of redshift in a real survey.
On the left side of Figure~\ref{fig:F_parameter_problems} we show the result of this exercise. In particular, we show the relative error made by assuming the Macquart parameterization with respect to the formal variance, calculated directly from the electron power spectrum, as a function of redshift. Additionally, we color-code the dependence on the feedback strength. If the $F$-parameter relation is an accurate representation of $p(\mathrm{DM}_\mathrm{cosmic}|z)$, all lines should coincide with zero. However, we can see that, depending on the feedback strength and the redshift considered, the $F$-parameter misestimates the variance by 20-40\%. The $F$-parameter does not capture the complex scaling of the variance with redshift, especially in the presence of feedback which itself evolves with redshift. Furthermore, the definition of the $F$-parameter ignores the fact that it should depend on the cosmological parameters which are not measured directly in the mean relation, $\langle\mathrm{DM}_\mathrm{cosmic}(z)\rangle$, for example the amplitude of fluctuations, $\sigma_8$, or any other parameter which is associated with perturbations and not the background expansion of the Universe. 

On the right side of Figure~\ref{fig:F_parameter_problems}, we now propagate the bias due to the $F$-parameter into an error of the matter power spectrum at the fiducial value $\log_{10}T_\mathrm{AGN} = 7.8$. To this end, we calculate:
\begin{equation}
\label{eq:response}
  \frac{\Delta P_\mathrm{mm}}{P_\mathrm{mm}} = \frac{\partial\log P_\mathrm{mm}}{\partial \log_{10}T_\mathrm{AGN}}
 \frac{\partial \log_{10}T_\mathrm{AGN}}{\partial \sigma_\mathrm{DM}(\Delta)}\delta\sigma_\mathrm{DM}(\Delta), 
\end{equation}
where we obtain $\delta\sigma_\mathrm{DM}(\Delta)$ from the left side in Figure~\ref{fig:F_parameter_problems}. The derivatives can be computed directly since the emulator is fully differentiable. One can see that, as expected, there are no effects on large scales, as feedback is not important there. On small scales, where upcoming weak lensing surveys obtain most of their signal, the assumption of the $F$-parameter generates 5-10\% errors in the matter power spectrum, well above the required accuracy for those surveys. 

This discussion highlights the shortcomings of the $F$-parameter. Firstly, it is highly degenerate with other cosmological parameters~\citep{2024ApJ...965...57B} and therefore is not a good summary statistic of feedback. Secondly, given the fact that the redshift scaling does not correspond to the formal variance of the electron field, biases are introduced into the estimation of the feedback strength, even if all cosmological parameters are perfectly known. We therefore conclude that the $F$-parameter is not a reliable choice for FRB analyses aimed at constraining baryonic feedback or cosmological parameters.

\subsection{Log-normal Parameterization} \label{subsec:lognormal_parameterization}

Given that the first and second moments of the $p(\mathrm{DM}_\mathrm{cosmic}|z)$ distribution can be analytically determined, we proceed by evaluating the effectiveness of fitting the $p(\mathrm{DM}_\mathrm{cosmic}|z)$ in simulations using a log-normal distribution which is fully specified by its aforementioned mean and variance. The log-normal distribution is known to provide a better approximation for the evolved matter density field than a Gaussian~\citep{2011A&A...536A..85H, 2016MNRAS.459.3693X, 2017MNRAS.466.1444C, 2017PhRvD..96l3510B}.

\begin{figure*}
\centering
\includegraphics[width=\textwidth]{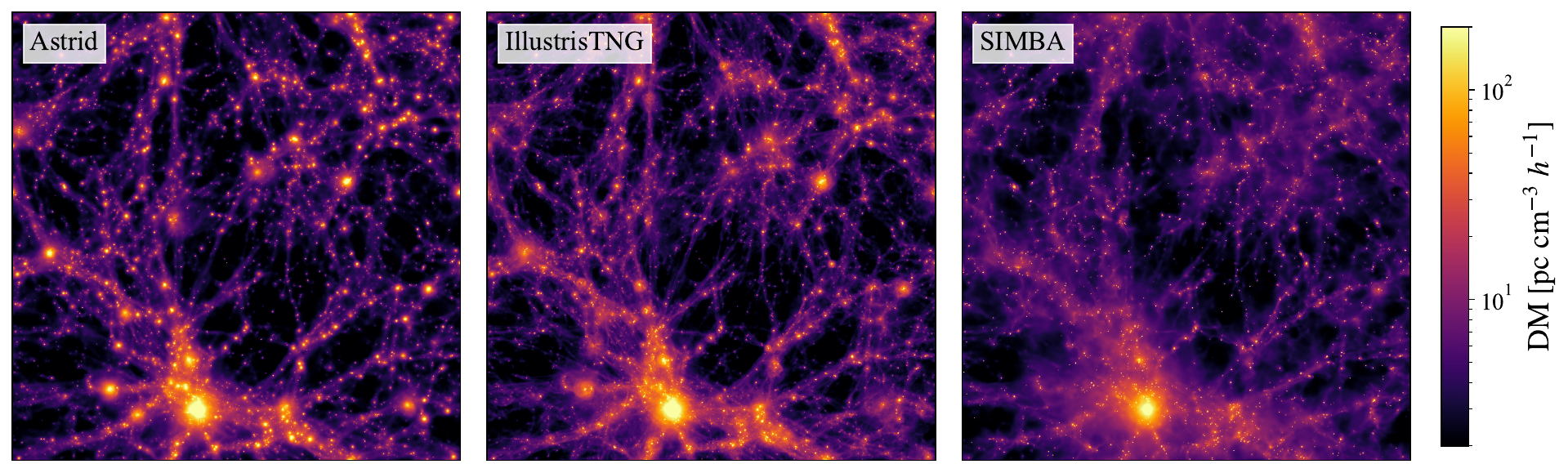}
\caption{The dispersion measure (DM) maps of the redshift $z = 0.00$ snapshots for the fiducial runs of \texttt{Astrid} (first column), \texttt{IllustrisTNG} (second column), and \texttt{SIMBA} (third column) simulation suites in the CAMELS project~\citep{2023ApJS..265...54V, 2023ApJ...959..136N}. Each simulation box has a comoving length of $25\,h^{-1}$~Mpc. Bright spots indicate the locations of halos, with variations in feedback prescriptions and subgrid model implementations across these simulations leading to different baryon distributions around them. \texttt{SIMBA} exhibits the strongest feedback, resulting in a smoother baryon distribution, while \texttt{Astrid} demonstrates the weakest feedback, leading to a more concentrated baryon distribution around halos.
}
\label{fig:CAMELS_simulations}
\end{figure*}

The variance of the density field increases with time due to gravitational clustering. However, the density field is constrained by a physical lower bound: the density cannot be negative, i.e., $\rho \geq 0$. This results in a skewed distribution. Voids can only reach a minimum relative under-density of $\delta = -1$. Conversely, over-dense regions, such as galaxy clusters, can have $\delta \gg 1$, leading to a long positive tail in the density distribution. Since the log-normal distribution naturally accommodates this asymmetry by ensuring positivity of density values while allowing a long-tailed distribution for over-densities, it serves as a reasonable approximation that works well under many circumstances. For example, \citet{2011A&A...536A..85H} and \citet{2017MNRAS.466.1444C} demonstrate that the log-normal model successfully describes the probability distribution of the weak lensing convergence ($\kappa_\mathrm{WL}$), a quantity similar to the DM by construction: $\kappa_\mathrm{WL}(\theta) = \int \mathrm{d}\chi~W(\chi)~\delta_\mathrm{m}(\chi, \theta)$, where $\theta$ represents the angular position, $W(\chi)$ is a redshift distribution and geometry dependent lensing efficiency or weight function (in analogy to $W_\mathrm{DM}$ in Equation~\ref{eqn:meanDM}) and $\delta_\mathrm{m}$ is the matter over-density field. The fits to the log-normal distribution parameterization are depicted as dotted lines in Figure~\ref{fig:macquart_lognormal_distribution}. The log-normal distribution offers a more accurate representation than those derived from the F-parameter functional form.

One might expect the central limit theorem (CLT) to drive the integrated density field (such as DM) towards Gaussianity, but two primary reasons prevent this. First, the variance of the density field evolves with time, meaning we are not summing identically distributed random variables. Second, density fluctuations at nearby redshifts are correlated due to gravitational interactions, violating the independence assumption of the CLT. Thus, the usual argument for Gaussianity does not hold in this context.

\newpage
\section{Feedback and Variance in Dispersion Measure in hydrodynamical simulations} \label{sec:feedback_sims}

In order to compute the variance in $\mathrm{DM}_\mathrm{cosmic}$ using the $P_\mathrm{ee} (k,z)$, we require a model that computes $P_\mathrm{ee} (k,z)$ as a function of a few physical parameters. Owing to the limitations of existing halo models in computing the gas power spectrum at relevant physical scales (see Appendix~\ref{subsec:HMcode}), we develop our own calibration dataset that maps cosmological and astrophysical feedback parameters in hydrodynamical simulations to the $P_\mathrm{ee} (k,z)$. This is enabled by the correlations of these parameters with the clustering statistics of baryons (see \S\ref{subsec:feedback_params}).  Given this calibration set, we can then interpolate the $P_\mathrm{ee} (k,z)$ as a function of these parameters when conducting the $\mathrm{DM}_\mathrm{exgal}-z$ analysis (see \S\ref{sec:simulation_based_inference}).

\subsection{CAMELS Project} \label{subsec:CAMELS}

The Cosmology and Astrophysics with MachinE Learning Simulations \citep[CAMELS;][]{2023ApJS..265...54V, 2023ApJ...959..136N} project constitutes an extensive suite of hydrodynamical simulations conducted in a periodic box of comoving size $25\,h^{-1}$Mpc, along with their gravity-only counterparts that have no feedback. The publicly available CAMELS dataset encompasses \texttt{IllustrisTNG}, \texttt{Astrid} and \texttt{SIMBA} simulation suites, which differ primarily in their subgrid implementation of feedback physics. Each of these simulation suites include a ``fiducial run'' with fixed parameters matching those of the original simulations. The differences in the strength of implemented feedback processes and baryon coupling efficiency in the fiducial run of these simulations is demonstrated in Figure~\ref{fig:CAMELS_simulations}, where the distinction in the distribution of baryons around halos is notable. 

In addition to the fiducial run, each simulation suite also includes variations of cosmological and astrophysical feedback parameters that spans a broad range of the parameter space. The stellar ($A_\mathrm{SN1}$ and $A_\mathrm{SN2}$) and AGN ($A_\mathrm{AGN1}$ and $A_\mathrm{AGN2}$) feedback parameters have been shown to correlate with the baryon distribution around halos~\citep{2024ApJ...967...32M, 2024arXiv240308611T}, thus we exclusively focus on variations of these four feedback parameters. The specific interpretation of these feedback parameters differs across simulation suites. For example, in \texttt{Astrid} and \texttt{IllustrisTNG}, $A_\mathrm{SN1}$ defines the galactic wind energy per unit star formation rate, whereas in \texttt{SIMBA}, it defines the galactic wind mass loading factor. 

The two CAMELS datasets employed in this study are: (i) the one-parameter-at-a-time ($1P$) dataset, which enables an isolated examination of the effects of individual feedback parameters on the key observables; and (ii) the Latin hypercube ($LH$) dataset, which provides a calibration dataset sampled across the multidimensional parameter space using a Latin hypercube design, thereby facilitating interpolation of the observables as a function of the underlying parameters. The $1P$ dataset of the \texttt{IllustrisTNG} simulations includes five distinct variations for each individual feedback parameter. Due to the degeneracies among these parameters (see \S\,\ref{subsec:caveats}) and their scale-dependent effects on the suppression of the matter power spectrum (see \S\,\ref{subsec:feedback_params}), it is necessary to compute the  $P_\mathrm{ee}(k,z)$ as a joint function of all relevant parameters. To this end, we construct our simulation-based model using the $LH$ dataset, which consists of 1000 realizations of the \texttt{IllustrisTNG} simulations wherein four feedback parameters and two cosmological parameters -- $\Omega_\mathrm{m}$ and $\sigma_8$ -- are simultaneously varied. In the following section, we describe the procedure for computing the power spectrum and correlation function for each of these datasets.

\begin{figure*}
\centering
\includegraphics[width=\textwidth]{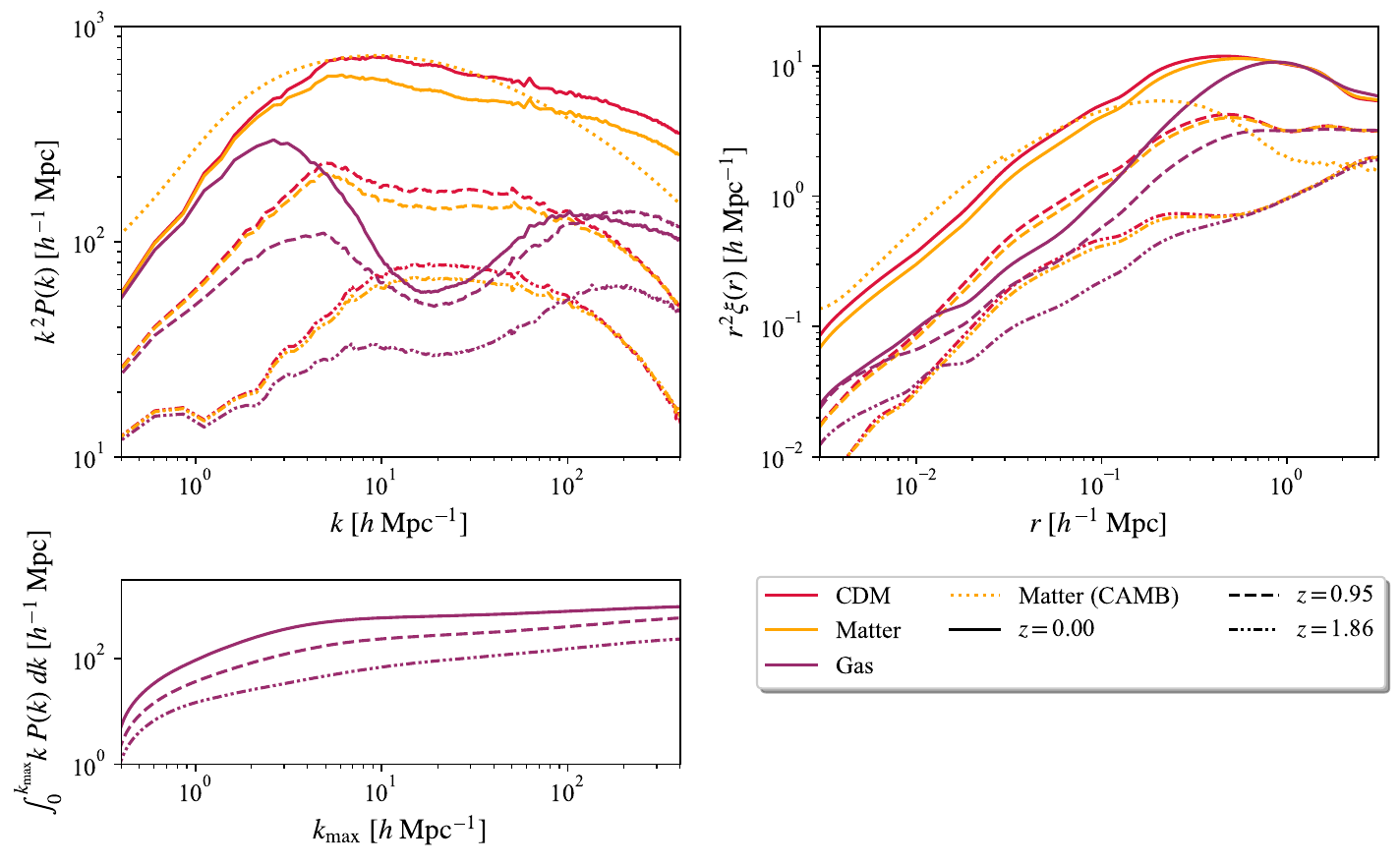}
\caption{The power spectra (first column) and correlation function (second column) for the matter, cold dark matter (CDM) and gas field components in the fiducial run of \texttt{IllustrisTNG} (normalized by the mean density of the corresponding component). The dotted line indicates the non-linear theory (using \citet{2021MNRAS.502.1401M} halo model) matter power spectrum and correlation function computed using \sw{CAMB}~\citep{2000ApJ...538..473L, 2002PhRvD..66j3511L, 2012JCAP...04..027H}. The bottom panel illustrates the integral of these physical quantities, highlighting that $\sim 95$\% of the contribution to DM variance comes from scales $\lesssim 10\,h\,\mathrm{Mpc}^{-1}$.}
\label{fig:power_spectra_correlation_fxn_gas_cdm}
\end{figure*}

\subsection{Power Spectrum and Correlation Function} \label{subsec:Pk_corrfunc}

We compute the power spectrum $P(k)$ and the correlation function $\xi(r)$ of the cold dark matter (CDM), gas and matter fields for all snapshots using \sw{Pylians}~\citep{Pylians}. The field over-densities are normalized by the mean density of the corresponding component. To accurately quantify the relevant scales contributing to the variance in DM$_\mathrm{cosmic}$, it is essential to construct the power spectrum across the full dynamic range of the simulations, spanning from the comoving box size of $25\,h^{-1}\text{Mpc}$ down to the gravitational softening scale of $2\,h^{-1}\text{kpc}$. This corresponds to a dynamic range of at least four orders of magnitudes. However, computing density contrast fields on grids with mesh sizes of $\mathcal{O}(10^4)$ is computationally prohibitive. To circumvent this limitation and extend the dynamic range, we employ the ``self-folding'' technique introduced by \citet{1998ApJ...499...20J}. The measurement of the correlation function in real-space circumvents any uncertainties due to discreteness effects. However, estimating the power spectrum from a finite dataset of random tracers introduces shot-noise, which we subtract from our raw power spectrum estimates. 

We show the resulting correlation functions and power spectra for the fiducial run of \texttt{IllustrisTNG} simulation in Figure~\ref{fig:power_spectra_correlation_fxn_gas_cdm} for the three fields considered here. Our estimates of the power spectra agree reasonably well with higher resolution simulations like TNG100 (75\,$h^{-1}$Mpc box, $2 \times 1820^3$ particles, 0.5\,$h^{-1}$kpc gravitational softening length) and TNG300 (205\,$h^{-1}$Mpc box, $2 \times 2500^3$ particles, 1\,$h^{-1}$kpc gravitational softening length), thus ensuring numerical convergence (see Figure~1 and 4 of \citet{2018MNRAS.475..676S} for comparison). Deviations from the theoretical non-linear~\citep{2021MNRAS.502.1401M} matter power spectrum~\citep[\sw{CAMB};][]{2000ApJ...538..473L, 2002PhRvD..66j3511L, 2012JCAP...04..027H} at large scales are apparent due to the limited volume of CAMELS simulation box. 

As expected, the gas field is less clustered than CDM at small scales ($k \gtrsim 1\,h\,$Mpc$^{-1}$) as gas is expelled from host halos by feedback. Consequently, gas does not trace CDM and the matter power spectrum is suppressed in hydrodynamical simulations with respect to their gravity-only counterpart. As we go to higher redshifts, the fractional suppression of the gas power spectrum with respect to the CDM power spectrum decreases since the gas (relatively) better traces matter at early times before the onset of feedback. Next, we use these gas power spectra to identify the scales that significantly contribute to the variance in DM$_\mathrm{cosmic}$.

\subsection{Relevant Scales for DM Variance} \label{subsec:relevant_scales}

The variance in DM$_\mathrm{cosmic}$ is governed by an integral over power spectrum (see Equation~\ref{eqn:variance}). In the bottom panel of Figure~\ref{fig:power_spectra_correlation_fxn_gas_cdm}, we assess whether these integrals converge within the range of scales accessible in the CAMELS project at several redshifts. Our analysis indicates that the integral converges by $\sim 95$\% ($\sim 99$\%) up to $ k_\mathrm{max} \sim 10\,h \,\mathrm{Mpc}^{-1}$ ($100\,h \,\mathrm{Mpc}^{-1}$), which is smaller than the comoving gravitational softening length of 2\,kpc.

\begin{figure*}[ht!]
\centering
\includegraphics[width=\textwidth]{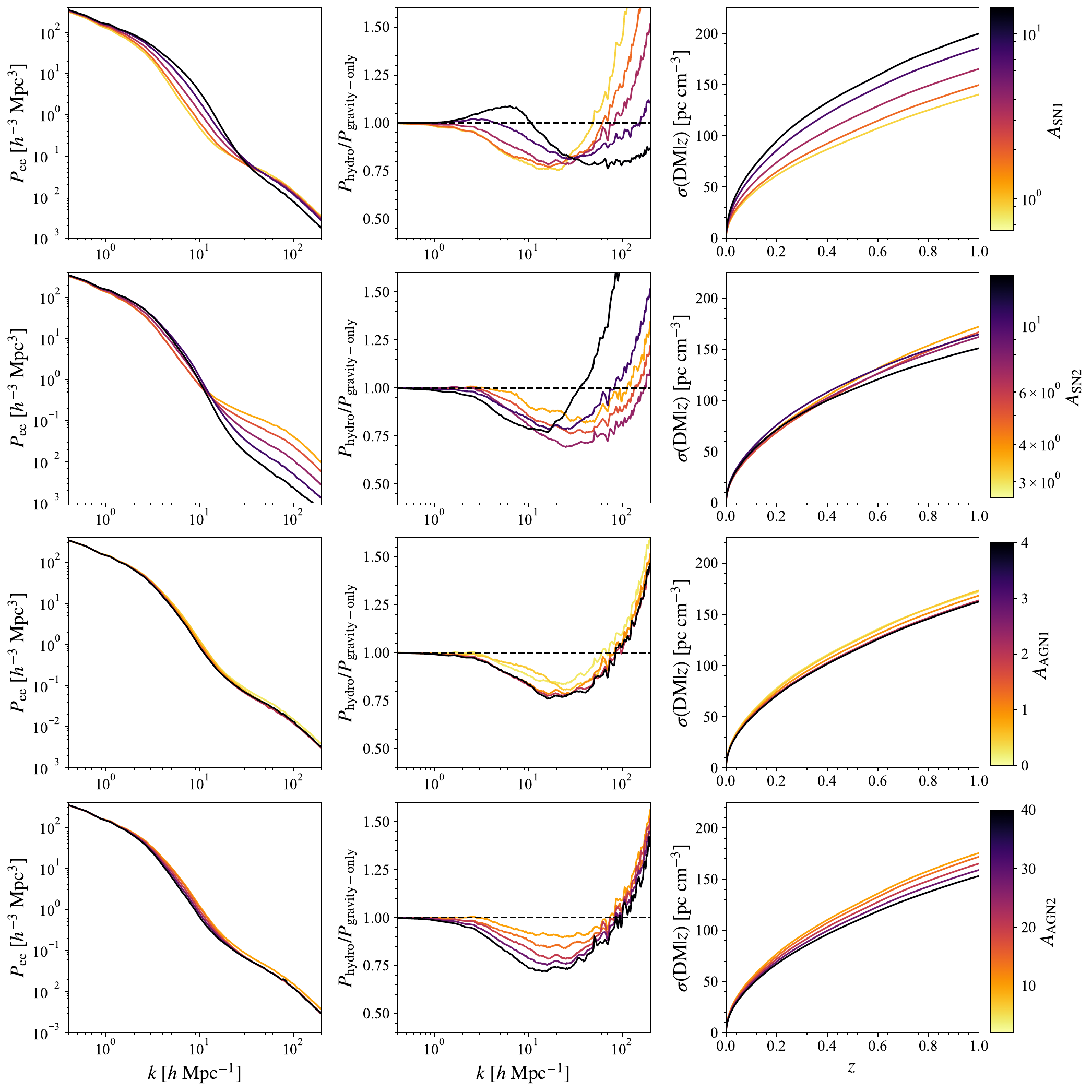}
\caption{The electron power spectrum (first column), suppression in the matter power spectrum (second column), and variance in $\mathrm{DM}_\mathrm{cosmic}$ (third column), as measured from the $1P$ dataset of simulations in \texttt{IllustrisTNG}, are analyzed as functions of the supernova and AGN feedback parameters. An increase in feedback energy enhances the suppression of the matter power spectrum and reduces the variance in $\mathrm{DM}_\mathrm{cosmic}$. The counterintuitive trend observed for $A_\mathrm{SN1}$ arises due to the complex interdependence of supernova and AGN feedback modes. Specifically, an efficient supernova feedback inhibits black hole growth and therefore, the threshold black hole mass for turning on AGN feedback is not reached.}
\label{fig:baryonic_feedback}
\end{figure*}

\begin{figure*}[ht!]
\centering
\includegraphics[width=\textwidth]{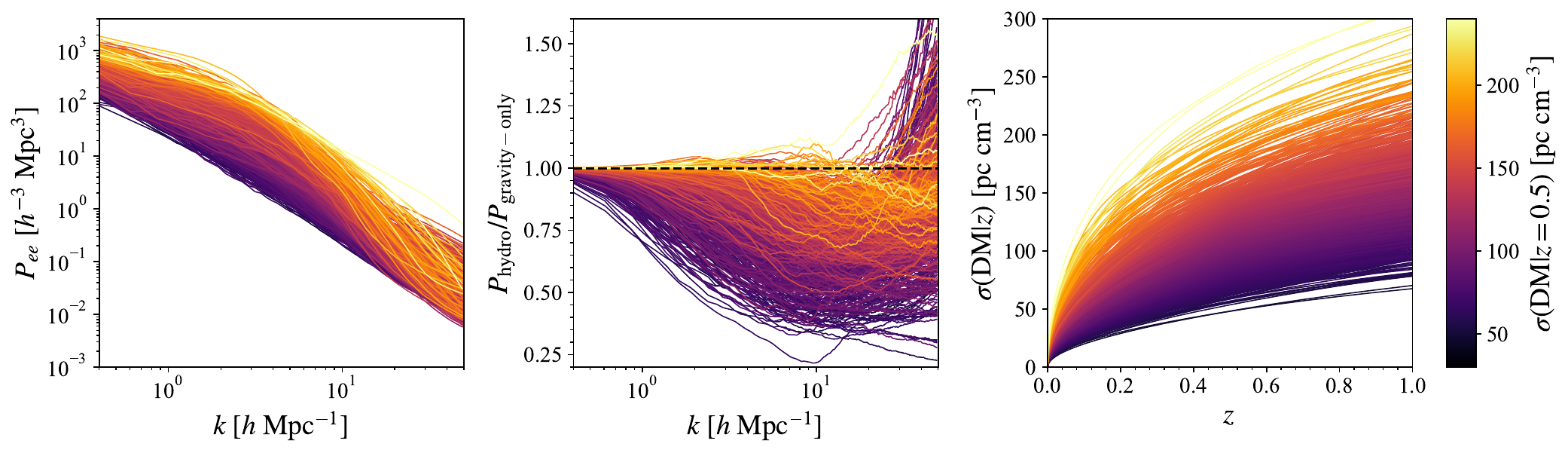}
\caption{The observed range of the electron power spectrum (top panel), suppression in the matter power spectrum (middle panel), and variance in $\mathrm{DM}_\mathrm{cosmic}$ (bottom panel) for the LH set, which comprises variations of the cosmological ($\Omega_\mathrm{m}, \sigma_8$) and astrophysical feedback ($A_\mathrm{SN1}, A_\mathrm{SN2}, A_\mathrm{AGN1}, A_\mathrm{AGN2}$) parameters in the \texttt{IllustrisTNG} simulation suite with priors listed in Table~\ref{table:par_summary}. The variations of \texttt{IllustrisTNG} range from negligible feedback strength (no suppression in the matter power spectrum and large variance in $\mathrm{DM}_\mathrm{cosmic}$) to strong feedback (suppressed to $25$\% at $k \sim 10\,h$Mpc$^{-1}$ and small variance in $\mathrm{DM}_\mathrm{cosmic}$), thereby highlighting the wide range of feedback scenarios that can be modeled using this calibration set.}
\label{fig:baryonic_feedback_LH_TNG}
\end{figure*}

\subsection{Sensitivity to Baryonic Feedback} \label{subsec:feedback_params}

We use the $1P$ dataset of \texttt{IllustrisTNG} simulations from the CAMELS project to investigate how baryonic feedback parameters, $\{A_\mathrm{SN1}, A_\mathrm{SN2}, A_\mathrm{AGN1}, A_\mathrm{AGN2}\}$, individually affect three key physical quantities: (i) the electron power spectrum, $P_\mathrm{ee}(k,z)$ at scales relevant to the variance in $\mathrm{DM}_\mathrm{cosmic}$, i.e., scales up to $k \sim 10\,h\,$Mpc$^{-1}$; (ii) the suppression of the matter power spectrum relative to gravity-only simulations, $P_\mathrm{hydro}/P_\mathrm{gravity-only}$; and (iii) the redshift evolution of the variance in $\mathrm{DM}_\mathrm{cosmic}$, $\sigma (\mathrm{DM}_\mathrm{cosmic}|z)$. These dependencies are illustrated in Figure~\ref{fig:baryonic_feedback}.

Figure~\ref{fig:baryonic_feedback} reveals three key trends. First, increasing the values of the baryonic feedback parameters -- and consequently the cumulative energy injected by feedback processes -- leads to a suppression of power in $P_\mathrm{ee}(k)$, a monotonic increase in the amplitude of the suppression $P_\mathrm{hydro}/P_\mathrm{gravity-only}$, and a monotonic decrease in $\sigma(\mathrm{DM}_\mathrm{cosmic}|z)$. A notable exception is the counterintuitive behavior of $A_\mathrm{SN1}$, which arises from the interplay between different feedback mechanisms. Specifically, increasing stellar feedback efficiency suppresses the early growth of seed black holes, which in turn delays AGN feedback activation in massive halos. This interplay is quantitatively reflected in the correlation matrix (Figure~\ref{fig:correlation_matrix}), where {$A_\mathrm{SN2}, A_\mathrm{AGN1}, A_\mathrm{AGN2}$} show positive correlations among themselves, but exhibit negative correlations with $A_\mathrm{SN1}$.

Second, the scale-dependence of feedback effects is evident: supernova feedback parameters primarily influence small-scale power, while AGN feedback exerts a more uniform suppression across all scales. Third, both $P_\mathrm{hydro}/P_\mathrm{gravity-only}$ and $\sigma(\mathrm{DM}\mathrm{cosmic}|z)$ are strongly anti-correlated with the feedback parameters -- except for $A_\mathrm{SN1}$, which, due to the delayed AGN activity it induces, is positively correlated with $\sigma(\mathrm{DM}_\mathrm{cosmic}|z)$.

\section{Simulation-based Model} \label{sec:simulation_based_inference}

In this section, we develop our hydrodynamical simulation-based model for inferring the FRB dispersion measure -- redshift relation. The model is calibrated using the variations of \texttt{IllustrisTNG} in the $LH$ dataset (see \S\,\ref{subsec:model_construction}) and validated on mock $\mathrm{DM}_\mathrm{cosmic}$ samples generated from the analytical $p(\mathrm{DM}_\mathrm{cosmic}|z)$ distribution in the fiducial run of \texttt{IllustrisTNG} (see \S\,\ref{subsec:mock_frb_generation}). We construct our likelihood and present forecasts for the expected constraints on the cosmological parameters and the $P_\mathrm{hydro}/P_\mathrm{gravity-only}$ by conducting an MCMC analysis (see \S\,\ref{subsec:likelihood_analysis}). We choose to do MCMC analysis instead of Fisher forecasts to avoid assumptions of linearity or Gaussianity in likelihood construction and to allow us to impose meaningful priors on the parameters of interest. We also test the ability of our \texttt{IllustrisTNG}-calibrated model to recover the $P_\mathrm{hydro}/P_\mathrm{gravity-only}$ in simulations with different subgrid feedback physics (see \S\ref{subsec:robustness}). Finally, we discuss the limitations and potential caveats associated with our approach (see \S\ref{subsec:caveats}).

\subsection{Model Construction}\label{subsec:model_construction}

We model $P_\mathrm{ee}(k,z)$ and $P_\mathrm{hydro}/P_\mathrm{gravity-only}$ using $LH$ dataset of \texttt{IllustrisTNG} simulation which varies two cosmological parameters and four astrophysical feedback parameters in the following ranges: $\Omega_\mathrm{m} \in (0.1, 0.5)$, $\sigma_8 \in (0.6, 1)$, $\log A_\mathrm{SN1} \in (-0.6, 0.6)$, $\log A_\mathrm{SN2} \in (-0.3, 0.3)$, $\log A_\mathrm{AGN1} \in (-0.6, 0.6)$ and $\log A_\mathrm{AGN2} \in (-0.3, 0.3)$. The range of feedback implementations in this dataset is visualized in Figure~\ref{fig:baryonic_feedback_LH_TNG}, where it varies from negligible feedback scenario with no suppression of matter power spectrum across all scales to strong feedback scenario with suppression of matter power spectrum to 25\% at small scales. The corresponding variance in DM$_\mathrm{cosmic}$ ranges from $\sigma (\mathrm{DM}_\mathrm{cosmic} | z = 0.5) \sim 250$\,pc\,cm$^{-3}$ in no feedback scenario to $\sigma (\mathrm{DM}_\mathrm{cosmic} | z = 0.5) \sim 50$\,pc\,cm$^{-3}$ in the strong feedback scenario. We perform nearest-neighbor interpolation on this calibration dataset to model $P_\mathrm{ee}(k,z)$ and $P_\mathrm{hydro}/P_\mathrm{gravity-only}$ as a function of these six parameters. We avoid doing linear interpolation since the number density of interpolation points in the 6-dimensional parameter space is quite low (see \S~\ref{subsec:caveats} for a more detailed discussion).

Our approach is in principle similar to \citet{2023MNRAS.523.2247S}, where the authors demonstrate a strong correlation between the $P_\mathrm{hydro}/P_\mathrm{gravity-only}$ and baryon fraction in galaxy groups with masses in the range of $M \sim 10^{13}-10^{14}\,M_\odot$. Parallel to our methodology, the authors developed a model mapping the baryon fraction in massive halos to the $P_\mathrm{hydro}/P_\mathrm{gravity-only}$, which was fitted to the ANTILLES suite of 400 hydrodynamical simulations spanning a broad range of baryonic feedback scenarios by varying stellar and AGN feedback parameters. Mirroring our approach, the $P_\mathrm{hydro}/P_\mathrm{gravity-only}$ in \texttt{HMcode}~\citep{2020A&A...641A.130M, 2021MNRAS.502.1401M} was fitted to the \texttt{BAHAMAS} suite of simulations. Therefore, building on these insights, our parameterization of $P_\mathrm{ee}(k,z)$ provides a physically motivated framework for capturing the impact of baryonic feedback, which is analogous to previous efforts in modeling $P_\mathrm{hydro}/P_\mathrm{gravity-only}$.

\begin{figure*}
\centering
\includegraphics[width=\textwidth]{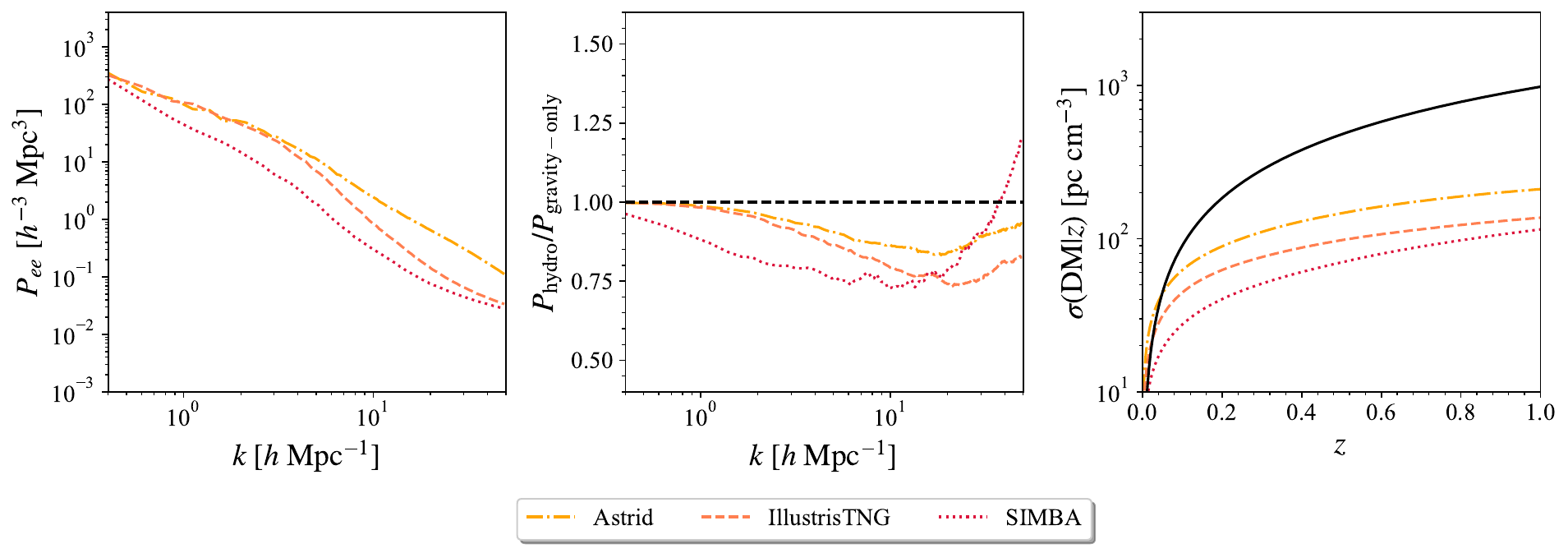}
\caption{The electron power spectrum (left panel), suppression in the matter power spectrum (middle panel), and variance in $\mathrm{DM}_\mathrm{cosmic}$ (right panel) examined across the \texttt{IllustrisTNG} (dashed line), \texttt{SIMBA} (dotted line), and \texttt{Astrid} (dash-dotted line) simulation suites for their fiducial runs. For reference, the mean FRB DM is depicted by the solid line. A stronger feedback implementation results in a more pronounced suppression of both the electron power spectrum and the matter power spectrum, while also leading to a lower variance in $\mathrm{DM}_\mathrm{cosmic}$. We show their measurements using our MCMC analyses in Figure~\ref{fig:suppression_varDM_results}.}
\label{fig:baryonic_feedback_all_sims}
\end{figure*}

\begin{figure*}
\centering
\includegraphics[width=\textwidth]{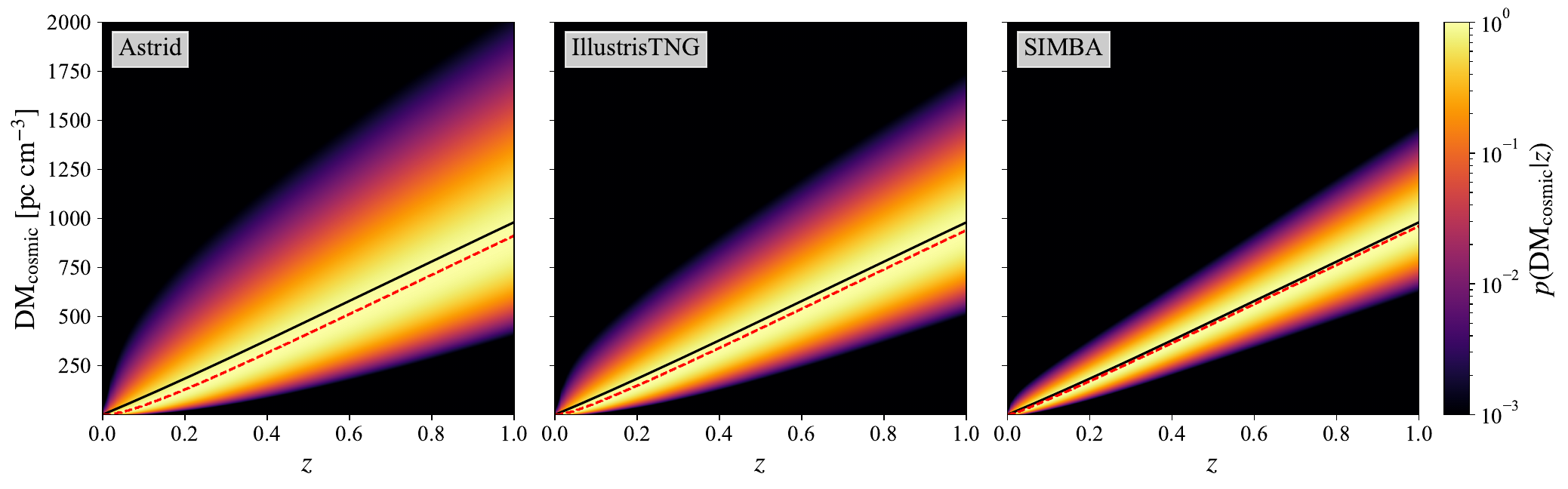}
\caption{The normalized $p(\mathrm{DM}_\mathrm{cosmic} | z)$ distribution for the fiducial runs of \texttt{Astrid} (first column), \texttt{IllustrisTNG} (second column) and \texttt{SIMBA} (third column) simulations. The mean (solid line) and mode (dashed line) of the distributions are shown for reference. While the mean of the $p(\mathrm{DM}_\mathrm{cosmic} | z)$ distribution remains the same across all simulations in their fiducial runs due to fixed cosmology, the mode deviates from the mean depending on the strength of the feedback. The simulated FRB population is sampled from these $p(\mathrm{DM}_\mathrm{cosmic} | z)$ distributions.}
\label{fig:DM_z_sims}
\end{figure*}

\subsection{Mock FRB Samples Generation}\label{subsec:mock_frb_generation}

Our goal is to validate our model on \texttt{IllustrisTNG} and test its ability to recover the $P_\mathrm{hydro}/P_\mathrm{gravity-only}$ in different simulations. To this end, we first need to generate mock samples of DM$_\mathrm{cosmic}$ in the fiducial run of all the three simulations under consideration -- \texttt{Astrid}, \texttt{IllustrisTNG} and \texttt{SIMBA}. We qualitatively demonstrated the widely different feedback strengths in these simulations in Figure~\ref{fig:CAMELS_simulations}. Quantitatively, these effects are imprinted on the clustering statistics of baryons and hence, the variance in DM$_\mathrm{cosmic}$. In Figure~\ref{fig:baryonic_feedback_all_sims}, we present the $P_\mathrm{ee}(k,z)$ and the $P_\mathrm{hydro}/P_\mathrm{gravity-only}$ for these simulations. Notably, \texttt{Astrid} and \texttt{SIMBA} represent two distinct extremes relative to \texttt{IllustrisTNG}, exhibiting weaker and stronger feedback implementations, respectively, thus impacting the power spectra at all scales. As a result, the variance in $\mathrm{DM}_\mathrm{cosmic}$ is consistently higher in \texttt{Astrid} and lower in \texttt{SIMBA} compared to \texttt{IllustrisTNG} across all redshifts.

\setlength{\tabcolsep}{4pt}
\begin{table*}
    \centering
    \begin{tabular}{lllllll}
        \toprule
        Parameter/ & Description & Prior & Fiducial & Posterior & Posterior & Posterior \\
        Observable & & & & $N_\mathrm{FRB} = 10^2$ & $N_\mathrm{FRB} = 10^3$ & $N_\mathrm{FRB} = 10^4$ \\
        \hline

        $H_0$ & Hubble constant & $[60, 80]$ & 67.7 & $71.18^{+4.17}_{-3.90}$ &  $67.47^{+2.70}_{-1.87}$ & $68.00^{+1.49}_{-0.55}$ \\
        
        $\Omega_\mathrm{m}$ & Present epoch matter density & $[0.1, 0.5]$ & 0.3 & $0.31^{+0.06}_{-0.07}$ & $0.33^{+0.04}_{-0.06}$ & $0.31^{+0.02}_{-0.05}$ \\
        
        $\sigma_8$ & RMS of matter density fluctuations & $[0.6, 1]$ & 0.8 & $0.79^{+0.07}_{-0.06}$ & $0.77^{+0.07}_{-0.05}$ & $0.78^{+0.06}_{-0.03}$ \\
        
        $\mu_\mathrm{host}$ & $p(\mathrm{DM}_\mathrm{host})$ parameter & $[4, 6]$ & 5 & $5.04^{+0.14}_{-0.16}$ & $5.00^{+0.05}_{-0.04}$ & $5.00^{+0.03}_{-0.01}$ \\
        
        $\sigma_\mathrm{host}$ & $p(\mathrm{DM}_\mathrm{host})$ parameter & $[0.2, 1]$ & 0.5 & $0.61^{+0.12}_{-0.12}$ & $0.52^{+0.03}_{-0.03}$ & $0.49^{+0.01}_{-0.01}$ \\
        
        $\log A_\mathrm{SN1}$ & Galactic wind energy per unit SFR & $[-0.6, 0.6]$ & 0 & $-0.17_{-0.30}^{+0.40}$ & $-0.18_{-0.27}^{+0.36}$ & $0.02_{-0.20}^{+0.07}$ \\
        
        $\log A_\mathrm{SN2}$ & Galactic wind speed & $[-0.3, 0.3]$ & 0 & $0.04_{-0.22}^{+0.18}$ & $-0.02_{-0.21}^{+0.20}$ & $-0.10_{-0.15}^{+0.03}$ \\
        
        $\log A_\mathrm{AGN1}$ & Energy per unit black hole accretion rate & $[-0.6, 0.6]$ & 0 & $0.03_{-0.42}^{+0.40}$ & $0.07_{-0.48}^{+0.38}$ & $-0.09_{-0.01}^{+0.34}$ \\
        
        $\log A_\mathrm{AGN2}$ & Ejection speed/burstiness & $[-0.3, 0.3]$ & 0 & $0.06_{-0.21}^{+0.17}$ & $0.04_{-0.21}^{+0.16}$ & $0.15_{-0.08}^{+0.10}$ \\

        SP($k=0.5$) & $P_\mathrm{hydro}/P_\mathrm{gravity-only}$ at $k=0.5\,h$\,Mpc$^{-1}$ & [0.86, 1.00] & 1.00 & $0.993_{-0.011}^{+0.006}$ & $0.994_{-0.008}^{+0.004}$ & $0.998_{-0.006}^{+0.001}$ \\
        
        SP($k=1$) & $P_\mathrm{hydro}/P_\mathrm{gravity-only}$ at $k=1\,h$\,Mpc$^{-1}$ & [0.70, 1.01] & 0.98 &  $0.971_{-0.032}^{+0.023}$ & $0.977_{-0.020}^{+0.015}$ & $0.990_{-0.019}^{+0.001}$ \\
        
        SP($k=10$) & $P_\mathrm{hydro}/P_\mathrm{gravity-only}$ at $k=10\,h$\,Mpc$^{-1}$ & [0.22, 1.11] & 0.79 & $0.766_{-0.092}^{+0.067}$ & $0.792_{-0.057}^{+0.052}$ & $0.844_{-0.062}^{+0.001}$ \\
        
        SP($k=50$) & $P_\mathrm{hydro}/P_\mathrm{gravity-only}$ at $k=50\,h$\,Mpc$^{-1}$ & [0.23, 2.39] & 0.83 & $0.909_{-0.164}^{+0.189}$ & $0.909_{-0.100}^{+0.190}$ & $0.873_{-0.053}^{+0.036}$ \\

        \hline
    \end{tabular}
    \caption{Summary of the priors, fiducial values and 68\% constraints on the posterior for cosmological parameters, feedback parameters and $P_\mathrm{hydro}/P_\mathrm{gravity-only}$ at various scales for the fiducial run of \texttt{IllustrisTNG} computed using our MCMC analysis.}
    \label{table:par_summary}
\end{table*}

The DM$_\mathrm{cosmic} - z$ relations for the fiducial runs of these simulations, constructed using our methodology of analytically computing the $\langle \mathrm{DM}_\mathrm{cosmic} (z) \rangle$ and the $\sigma^2 [\mathrm{DM}_\mathrm{cosmic} (z)]$ of the $p(\mathrm{DM}_\mathrm{cosmic}|z)$ distribution with log-normal parameterization are shown in Figure~\ref{fig:DM_z_sims}. Since the cosmological parameters in all the simulations are the same, the $\langle \mathrm{DM}_\mathrm{cosmic} (z) \rangle$ of the $p(\mathrm{DM}_\mathrm{cosmic}|z)$ distribution is the same across the three simulation suites. However, due to different feedback implementations, the $\sigma [\mathrm{DM}_\mathrm{cosmic} (z)]$ decreases as the feedback strength across the simulations increases and hence, the deviation of the mode from the mean of the distributions reduces.

For our MCMC analysis, we generate mock samples of $\mathrm{DM}_\mathrm{cosmic}$ by sampling from the $p(\mathrm{DM}_\mathrm{cosmic}|z)$ distribution. We then compute the extragalactic dispersion measure as $\mathrm{DM}_{\mathrm{exgal}} = \mathrm{DM}_{\mathrm{cosmic}} + \mathrm{DM}_{\mathrm{host}}/(1+z)$, where $\mathrm{DM}_{\mathrm{host}}$ is sampled from a log-normal distribution with $\mu_{\mathrm{host}} = 5$ and $\sigma_{\mathrm{host}} = 0.5$~\citep{2024arXiv240916952C}. Practically, the FRB redshift distribution follows $z^2 e^{-\alpha z}$, where $\alpha$ parameterize the completeness of the survey. Since the goal of this work is to illustrate the new methodology, for simplicity we sample redshifts from a uniform distribution up to $z=1$ and assume that the redshift of each FRB source is known. Investigation of the impact of survey completeness is destined for future work, and therefore, these results should not be treated as accurate forecasts (see \S\,\ref{subsec:caveats}).

\subsection{Likelihood Analysis and Model Validation} \label{subsec:likelihood_analysis}

\begin{figure*}
\centering \includegraphics[width=\textwidth]{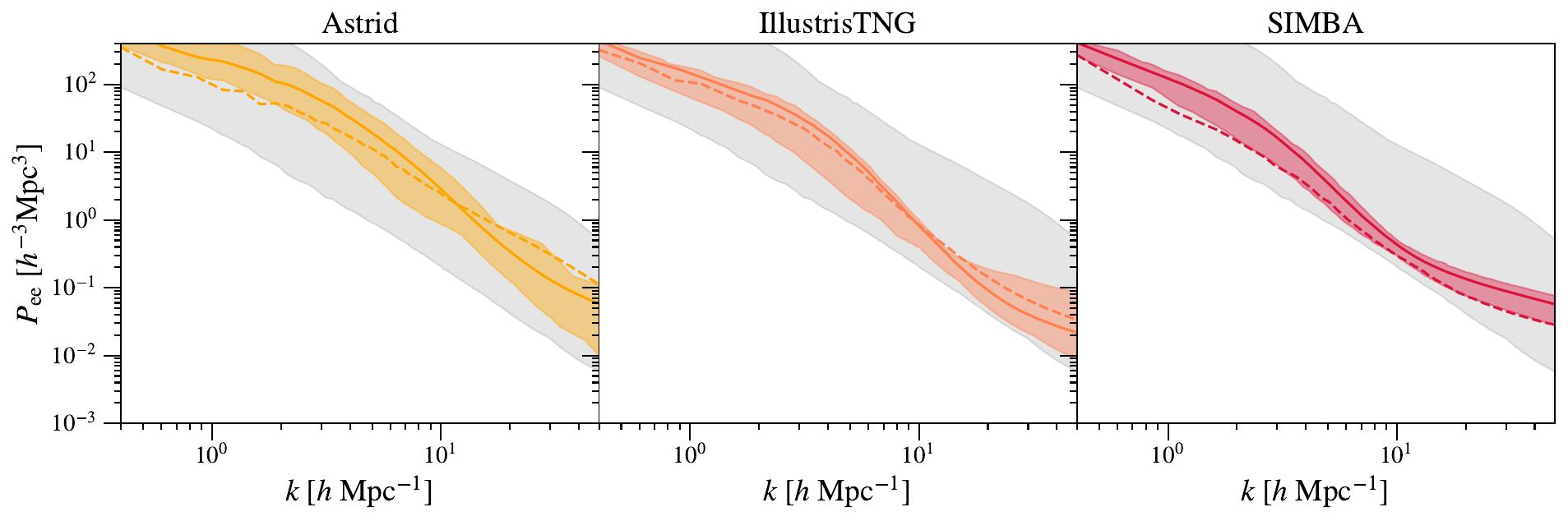}
\centering \includegraphics[width=0.99\textwidth]{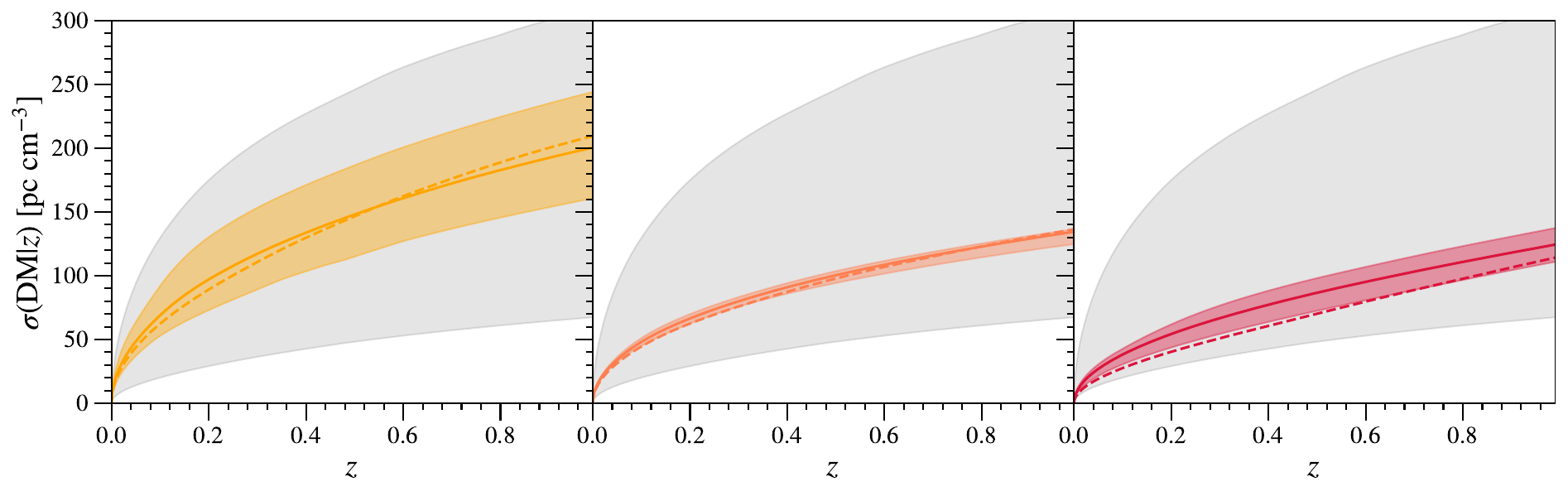}
\centering \includegraphics[width=0.99\textwidth]{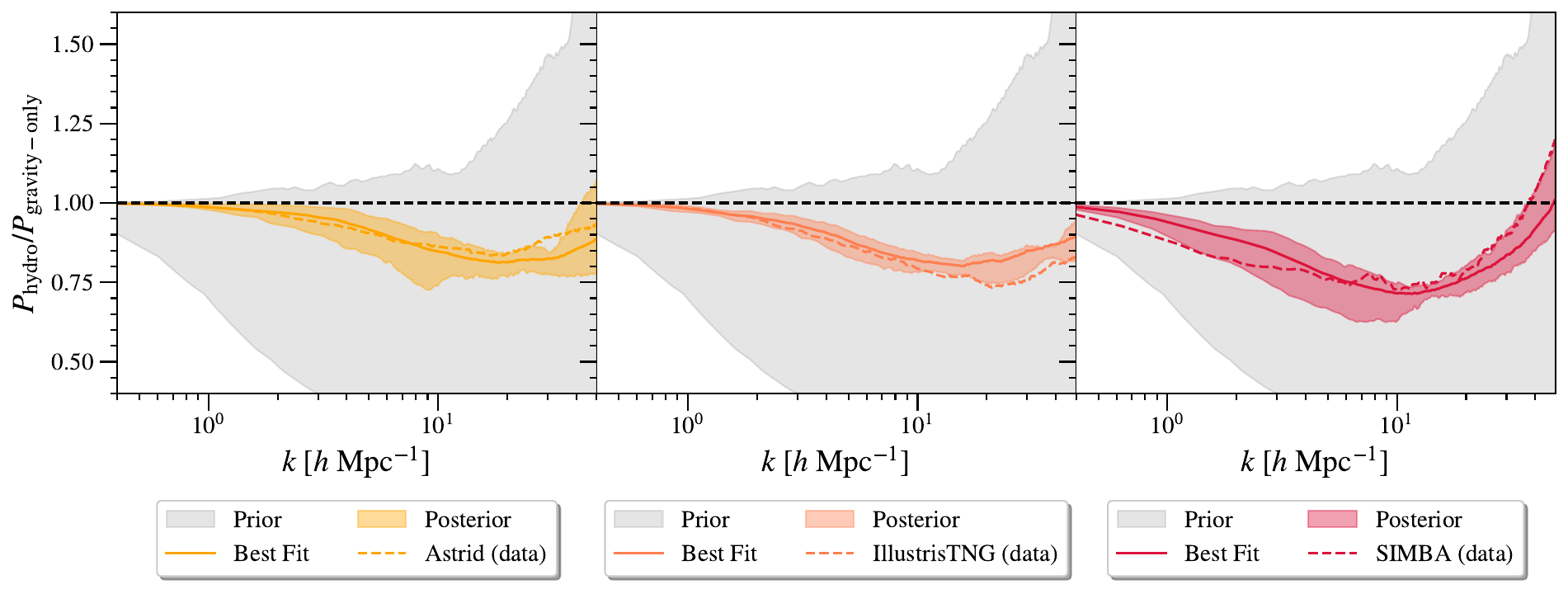}
\caption{Constraints on the electron power spectrum (top panels), variance in $\mathrm{DM}_\mathrm{cosmic}$ (middle panels) and suppression of the matter power spectrum (bottom panels) at redshift $z=0.00$ for the fiducial runs of the \texttt{Astrid} (first column), \texttt{IllustrisTNG} (second column), and \texttt{SIMBA} (third column) simulation suites. For reference, the prior range (gray region) on the cosmological and feedback parameters is shown. The true electron power spectrum, variance in $\mathrm{DM}_\mathrm{cosmic}$ and suppression in the matter power spectrum (dashed lines) are well recovered by the constrained posterior ($68$\% confidence interval, shaded region) across all simulation suites. The minor deviations observed at large scales primarily stem from the limited volume of the simulation box in CAMELS, as evidenced by the reduced amplitude of the matter power spectrum in comparison to the nonlinear theory predictions computed using \sw{CAMB} at large scales where feedback may not dominate.}
\label{fig:suppression_varDM_results}
\end{figure*}

For a log-normal parameterization, the distribution $p(\mathrm{DM}_\mathrm{cosmic} | z)$ is fully characterized by the mean $\langle \mathrm{DM}_\mathrm{cosmic} (z) \rangle$ and the standard deviation $\sigma [\mathrm{DM}_\mathrm{cosmic} (z)]$. Equation~\ref{eqn:meanDM} establishes that $\langle \mathrm{DM}_\mathrm{cosmic} (z) \rangle$ is proportional to $\Omega_\mathrm{b} H_0 f_\mathrm{d}$. The degeneracy between $f_\mathrm{d}$ and $\Omega_\mathrm{b} H_0$ can be effectively mitigated by incorporating priors on the baryon density confined within galaxies~\citep{2020Natur.581..391M}\footnote{The dominant source of uncertainty in $f_\mathrm{d}$ arises from errors in empirical stellar mass density estimates, which are contingent upon the assumed stellar initial mass function~\citep{2022MNRAS.516.4862J}.}. For the purpose of generating mock FRB DM samples and conducting MCMC analyses, we adopt a fixed value of $f_\mathrm{d} = 0.9$. The degeneracy between $\Omega_\mathrm{b}$ and $H_0$ can be broken by incorporating prior on $\Omega_\mathrm{b} H_0^2$ as inferred by primordial deuterium abundance measurements of near-pristine absorption systems and big bang nucleosynthesis (BBN) calculations\footnote{The BBN calculations use reaction cross-sections of various elements as inputs to compute the final abundances~\citep{2018ApJ...855..102C}. The uncertainty involved in the computation of baryon-to-photon ratio using the primordial deuterium abundance measurements is due to the systematic discrepancies between the theoretically and empirically measured $p(n,\gamma)d$, $d(p,\gamma)^3\mathrm{He}$ and $^3\mathrm{He}(\alpha,\gamma)^7\mathrm{Be}$ reaction cross-sections~\citep{2016ApJ...830..148C}.}. 

Marginalizing over $f_\mathrm{d}$ and $\Omega_\mathrm{b} H_0^2$ implies that the mean $\mathrm{DM}_\mathrm{cosmic}$ only depends on the cosmology as $\langle \mathrm{DM}_\mathrm{cosmic} (z) \rangle \propto H_0^{-1}$. On the other hand, the variance in $\mathrm{DM}_\mathrm{cosmic}$, defined in Equation~\ref{eqn:variance}, exhibits a complex dependence on cosmological parameters \{$H_0, \Omega_\mathrm{m}, \sigma_8$\} and feedback parameters $\{A_\mathrm{SN1}, A_\mathrm{SN2}, A_\mathrm{AGN1}, A_\mathrm{AGN2}\}$ used to define the $P_\mathrm{ee}(k,z)$. For each MCMC sample, the $P_\mathrm{ee}(k,z)$ is computed using nearest-neighbor interpolation on the \texttt{IllustrisTNG} $LH$ calibration dataset (see \S\,\ref{subsec:model_construction}). Therefore, given the $p(\mathrm{DM}_\mathrm{cosmic} | z)$ model parameters $\theta = \{H_0, \Omega_\mathrm{m}, \sigma_8, A_\mathrm{SN1}, A_\mathrm{SN2}, A_\mathrm{AGN1}, A_\mathrm{AGN2}\}$ and the rest-frame $p(\mathrm{DM}_\mathrm{host})$ log-normal distribution with parameters $\{\mu_\mathrm{host}, \sigma_\mathrm{host}\}$ (see Equation~\ref{eqn:DMhost}), the likelihood for a single FRB can be written as
\begin{equation}
    \begin{aligned}
        \mathcal{L}(\mathrm{DM}_\mathrm{exgal,i} | & z_{i}, \theta) = \int_0^{\mathrm{DM}_\mathrm{exgal,i}} 
        p(\mathrm{DM}_\mathrm{cosmic}|z_{i}, \theta) \\
        &\quad \times p\left( \mathrm{DM}_\mathrm{host} | z_{i}, \mu_\mathrm{host}, \sigma_\mathrm{host} \right) 
        \mathrm{d} \mathrm{DM}_\mathrm{cosmic}, \\
    \end{aligned}
\end{equation}
where $\mathrm{DM}_\mathrm{host} = (\mathrm{DM}_\mathrm{exgal,i} - \mathrm{DM}_\mathrm{cosmic}) \times (1+z_{i})$. Assuming that all the FRB observations are independent, the posterior for the parameters $\Gamma = \{\theta, \mu_\mathrm{host}, \sigma_\mathrm{host}\}$ is
\begin{equation}
    p(\Gamma | \mathrm{obs}) \propto \prod_{i=1}^{N_\mathrm{FRB}} \mathcal{L}(\mathrm{DM}_\mathrm{exgal,i} | z_i, \Gamma) \times \pi(\Gamma),
\end{equation}
where $N_\mathrm{FRB}$ is the number of FRB observations in our sample and $\pi(\Gamma)$ is the prior distribution on the free parameters. We assume that the priors are all independent and uniform (see Table~\ref{table:par_summary}). We sample the posteriors using the dynamic nested sampling routine \sw{dynesty}~\citep{Speagle_2020}.

We perform MCMC analysis for $10^2$, $10^3$ and $10^4$ FRB samples generated in the fiducial run of \texttt{IllustrisTNG} and list the 68\% confidence intervals of constrained parameters and the $P_\mathrm{hydro}/P_\mathrm{gravity-only}$ at various scales in Table~\ref{table:par_summary}. With a sample of $10^4~(10^3,~10^2)$ FRBs, we forecast 2\% (4\%,~6\%) constraint on $H_0$, 16\% (18\%,~23\%) constraint on $\Omega_\mathrm{m}$, 8\% (9\%,~10\%) constraint on $\sigma_8$, and 2\% (2\%, 3\%) constraints on host galaxy DM distribution parameters. All four feedback parameters are consistent with the true values within 3$\sigma$.

We illustrate our constraints with $10^4$ samples by comparing the true and the recovered $P_\mathrm{ee}(k,z)$, redshift evolution of the variance in DM$_\mathrm{cosmic}$ and the $P_\mathrm{hydro}/P_\mathrm{gravity-only}$ for \texttt{IllustrisTNG} in the second column of Figure~\ref{fig:suppression_varDM_results}. The gray shaded region denotes the range permitted by priors summarized in Table~\ref{table:par_summary} (see Figure~\ref{fig:baryonic_feedback_LH_TNG}). Given the feedback parameterization, the posterior distributions successfully capture the true $P_\mathrm{hydro}/P_\mathrm{gravity-only}$. The 68\% confidence interval of samples drawn from the posterior constrain the $P_\mathrm{hydro}/P_\mathrm{gravity-only}$ within 0.6\%, 2\%, 8\% and 15\% for scales $k=0.5,\,1,\,10$ and $50\,h\,$Mpc$^{-1}$, respectively. Given the current state of the field, with $10^2$ FRBs, the $P_\mathrm{hydro}/P_\mathrm{gravity-only}$ can be constrained within 1\%, 3\%, 12\% and 20\% for scales $k=0.5,\,1,\,10$ and $50\,h\,$Mpc$^{-1}$, respectively. For reference, the prior spans the $P_\mathrm{hydro}/P_\mathrm{gravity-only}$ in the range of $(0.86, 1.00)$, $(0.70, 1.01)$, $(0.22, 1.11)$ and $(0.23, 2.39)$ at scales of $k=0.5,\,1,\,10$ and $50\,h\,$Mpc$^{-1}$, respectively. This illustrates the ability of statistical FRB samples to constrain the $P_\mathrm{hydro}/P_\mathrm{gravity-only}$ with percent-level precision at large scales and $\sim$10\% precision at small scales.

\subsection{Robustness to Different Feedback Prescriptions} \label{subsec:robustness}

Having demonstrated the accuracy of our model on the \texttt{IllustrisTNG} simulation, we now assess its adaptability to alternative simulation suites, namely \texttt{Astrid} and \texttt{SIMBA}. This evaluation is crucial to ascertain the model's adaptability to capture diverse feedback scenarios that may be prevalent in the Universe. 

We perform our MCMC analysis using a sample of $10^4$ FRBs drawn from the fiducial runs of \texttt{Astrid} and \texttt{SIMBA} simulations. Since the four feedback parameters in all the simulations have different physical interpretation, we assess the accuracy of our inferred parameters by evaluating the resulting $P_\mathrm{hydro}/P_\mathrm{gravity-only}$ and the variance in $\mathrm{DM}_\mathrm{cosmic}$ as a function of redshift against the expected true values from the simulations in Figure~\ref{fig:suppression_varDM_results}. 

We find that our inference model, calibrated to the $LH$ dataset of \texttt{IllustrisTNG} simulation suite in CAMELS project, generalizes well to accurately recover the redshift evolution of the variance of DM$_\mathrm{cosmic}$ in \texttt{SIMBA} and \texttt{Astrid} simulation suites. Thus, our proposed methodology offers a substantial advantage over the $F$-parameterization, accommodating the redshift-dependent evolution of the variance of the $p(\mathrm{DM}_\mathrm{cosmic} | z)$ distribution by construction.

Our posterior distributions also successfully capture the true $P_\mathrm{hydro}/P_\mathrm{gravity-only}$ in \texttt{SIMBA} and \texttt{Astrid} simulations. The 68\% confidence interval of samples drawn from the posterior constrain the $P_\mathrm{hydro}/P_\mathrm{gravity-only}$ in \texttt{Astrid} within 0.5\%, 2\%, 18\% and 32\% at scales $k=0.5,\,1,\,10$ and $50\,h\,$Mpc$^{-1}$, respectively. Similarly, in \texttt{SIMBA}, the recovery accuracy is 2.5\%, 8\%, 16\% and 22\% at scales $k=0.5,\,1,\,10$ and $50\,h\,$Mpc$^{-1}$, respectively. For \texttt{SIMBA}, the accuracy of the fit deteriorates at large scales, a limitation we attribute to the relatively small box size of the CAMELS project. This effect is evident when comparing the matter power spectrum in simulations to the nonlinear theory predictions, where we observe an underestimation of power at large scales (see Figure~\ref{fig:power_spectra_correlation_fxn_gas_cdm}). Nevertheless, we have demonstrated that our methods are flexible for different feedback scenarios that may be prevalent in the Universe.

\subsection{Caveats} \label{subsec:caveats}

So far, we have developed and validated the performance of our \texttt{IllustrisTNG}-calibrated modeling framework. We have also demonstrated the effectiveness of our model in recovering the $P_\mathrm{hydro}/P_\mathrm{gravity-only}$ and the variance in DM$_\mathrm{cosmic}$ in various simulations. Nevertheless, there remain five key limitations of this work that should be addressed in future studies to facilitate application to real-world datasets and to accurately quantify systematic uncertainties arising from the modeling assumptions.

\subsubsection{Variance from Large Scales}

We determined that the maximum scale relevant for computing the variance of the FRB DM is $k_\mathrm{max} \sim 10\,h\,\mathrm{Mpc}^{-1}$. However, due to the limited box size of the CAMELS simulations, we are unable to accurately capture the variance contributed by large-scale structures. To illustrate this, we compare the scale-dependent integral of the matter power spectrum, $\int_{k_\mathrm{min}}^{k_\mathrm{max}} k P_\mathrm{mm}(k) \, \mathrm{d}k$, derived from hydrodynamical simulations with predictions from non-linear theory (see Figure~\ref{fig:power_spectra_correlation_fxn_gas_cdm}). Our analysis reveals that approximately 4\% and 10\% of the variance are unaccounted for when integrating down to $k_\mathrm{min} \sim 1\,h\,\mathrm{Mpc}^{-1}$ and $k_\mathrm{min} \sim 0.01\,h\,\mathrm{Mpc}^{-1}$, respectively. This highlights the importance of calibrating our model using a significantly larger simulation volume—preferably at least an order of magnitude greater—to accurately account for variance on large scales. Currently, the 50~cMpc simulation boxes in the CAMELS project are under works. Another practical solution could be to stitch the small scale power spectrum from CAMELS with linear theory predictions at large scales since the impact of baryonic feedback at large scales is negligible.

\begin{figure}
\centering
\includegraphics[width=\columnwidth]{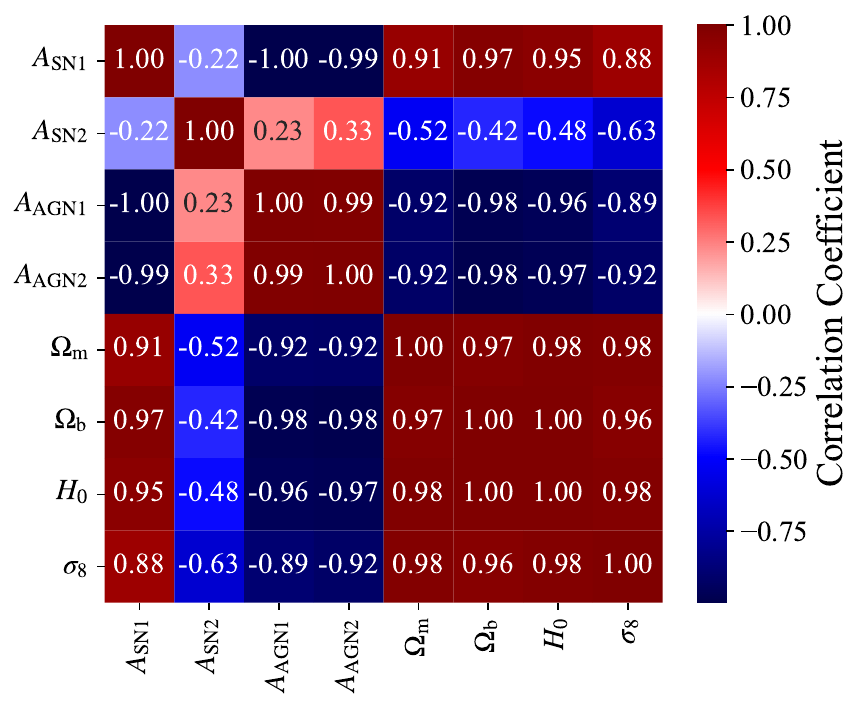}
\caption{The correlation among cosmological and feedback parameters that characterize the variance in $\mathrm{DM}_{\mathrm{cosmic}}$ is evident, thus supporting the necessity of parameterizing the variance in terms of physical quantities, such as the electron power spectrum, which directly encodes these parameters.}
\label{fig:correlation_matrix}
\end{figure}

\subsubsection{Parameter Degeneracy}

We note that $H_0$ and $\Omega_\mathrm{b}$ are degenerate with feedback parameters \{$A_\mathrm{SN1},A_\mathrm{SN2}, A_\mathrm{AGN1}, A_\mathrm{AGN2}$\} (the effects of cosmology on feedback itself are of higher order) and other cosmological parameters $\{\Omega_\mathrm{m}, \sigma_8\}$, suggesting that the $LH$ dataset calibration must be expanded to allow us to fit for them in the variance of DM$_\mathrm{cosmic}$. We use the $1P$ dataset of simulations to evaluate the degree of correlation of the impact of the four feedback parameters and four cosmological parameters on the variance ($\sigma^2(z)$) of DM$_\mathrm{cosmic}$ as the normalized dot product: 
\begin{equation}
    R_{ij} = \frac{{u_i \cdot u_j}}{{||u_i||~||v_j||}},
\end{equation} 
\begin{equation}
    u_i = \frac{\partial \log \sigma(\theta_i)}{\partial \theta_i} \approx \frac{\sigma(\theta_i + \Delta \theta_i) - \sigma(\theta_i)}{\sigma(\theta_i + \Delta \theta_i/2) \cdot \Delta \theta_i}.    
\end{equation} 
We show the correlation matrix thus computed in Figure~\ref{fig:correlation_matrix}. We observe that all cosmological parameters are positively correlated and these cosmological parameters are negatively correlated with all feedback parameters but $A_\mathrm{SN1}$. In other words, whatever makes feedback stronger will anti-correlate with the three cosmological parameters and vice-versa.

\subsubsection{Interpolation in Sparse Dataset}

Our work is based on $LH$ set, which comprises 1000 variations of \texttt{IllustrisTNG} in 6 dimensions. Interpolating in a six-dimensional space with only 1000 data points is generally inadequate due to the challenges posed by the curse of dimensionality. In high-dimensional spaces, the number of points required to maintain a consistent sampling density increases exponentially with the number of dimensions. Sparse sampling in high dimensions leads to interpolation errors, especially for methods like nearest-neighbor or linear interpolation that rely heavily on local data density. To improve interpolation accuracy under such sparsity constraints, one might consider dimensionality reduction techniques, like PCA~\citep{2018PhRvD..98d3532T}, or employing surrogate modeling approaches~\citep{2023JCAP...07..054D} or constructing a neural network-based emulator that can capture the underlying function more efficiently in high-dimensional spaces~\citep{2022JCAP...04..046N}.

\subsubsection{Observational Biases and Selection Effects}

We note that while the framework developed in this work is tested on mock FRBs sampled from hydrodynamical simulations, for observed FRBs in the real world, this framework may have to be modified to account for the impact of the interferometer detection efficiency and the FRB population characteristics (such as, the energetics/luminosity function) on the observed DM$_\mathrm{cosmic}-z$ relation and the constrained cosmological parameters~\citep{2022MNRAS.509.4775J, 2022MNRAS.516.4862J}. The detection efficiency is affected by DM, intrinsic burst width, scattering, DM smearing, observing resolution and telescope beam shape, and impact our ability to discover high DM events, which may lead to a biased inference of the tail of $p(\mathrm{DM}_\mathrm{cosmic}|z)$ at high redshifts.

\subsubsection{Likelihood Construction}

We know that the evolution of the density field through non-linear gravitational collapse leads to deviations from Gaussianity and the log-normal approximation of the integrated density field captures the skewness of the distribution and respect physical constraints, thus presenting it as a useful approximation, despite the absence of a rigorous theoretical derivation. Going forward, the likelihood-free inference approach of \citet{2024arXiv241007084K} can be used to quantify any deviations from log-normality. This approach enables posterior estimation without requiring an explicit likelihood function, instead relying on forward simulations of the large scale structure using \sw{GLASS}~\citep{2023OJAp....6E..11T} (which assumes a log-normal density field distribution) to evaluate the posterior.

\section{Conclusions} \label{sec:conclusions}

In this study, we have demonstrated the potential of FRBs in constraining the suppression of the matter power spectrum caused by baryonic feedback. This underscores the complementary role of FRBs alongside other probes of baryons. Owing to their ability to probe halos across a wide range of physical scales up to $k \sim 10\,h\,\mathrm{Mpc}^{-1}$, FRBs offer a promising avenue for precision cosmology. This capability aligns with the objectives of next-generation cosmology experiments focused on gravitational lensing analyses. We now provide a summary of the principal outcomes of our work.

\begin{itemize}[leftmargin=*]
    \item We discussed limitations of the widely-used method to parameterize the $p(\mathrm{DM}_\mathrm{cosmic}|z)$ distribution, which incorporates the impact of baryonic effects using the $F$-parameter. First, the $F$-parameter is defined under the assumption of Poisson statistics, which breaks down when filamentary structures on scales of hundreds of kiloparsecs significantly contribute to the variance in $\mathrm{DM}_\mathrm{cosmic}$. Second, the approximation $\sigma_\mathrm{DM} = F z^{-1/2}$ is invalid in the non-Euclidean limit at redshifts $z \gtrsim 0.5$ and the assumption of a constant $F$-parameter breaks down due to the evolution of the variance in $\mathrm{DM}_\mathrm{cosmic}$ with redshift~\citep{2024ApJ...965...57B}. Third, the $F$-parameter is degenerate with cosmological parameters, thus making it a suboptimal summary statistic of feedback. Fourth, it does not accurately capture the redshift evolution of the formal variance in DM, thus introducing biases into the estimates of feedback strength (see Figure~\ref{fig:F_parameter_problems}). Lastly, as shown in Figure~\ref{fig:macquart_lognormal_distribution}, the Equation~\ref{eqn:macquart_expression} presents an inaccurate fit to the $p(\mathrm{DM}_\mathrm{cosmic}|z)$ for FRB sightlines sampled from hydrodynamical simulations, thus motivating the need for a better parameterization.

    \item Since the closed form expression for the Fourier transform of the probability density function of DM depends on the unknown halo gas profiles~\citep{2007ApJ...671...14Z, 2014ApJ...780L..33M}, we propose to model $p(\mathrm{DM}_\mathrm{cosmic}|z)$ as a log-normal distribution which can be fully specified using the mean and the variance. These two moments can be analytically expressed in terms of cosmological parameters and the electron power spectrum $P_\mathrm{ee}(k,z)$ using first-principles (see Equation~\ref{eqn:meanDM} and \ref{eqn:variance}). We demonstrate that a log-normal formulation leads to an accurate fit to $p(\mathrm{DM}_\mathrm{cosmic}|z)$ for mock FRB samples in hydrodynamical simulations (see Figure~\ref{fig:macquart_lognormal_distribution}). Although there is no strict first-principles derivation proving that the log-normal distribution must describe the integrated density field, it serves as an empirically motivated approximation that works well under many circumstances, including the weak lensing convergence~\citep{2016MNRAS.459.3693X, 2017MNRAS.466.1444C, 2017PhRvD..96l3510B}.

    \item Due to the lack of accurate small-scale gas power spectrum halo models ($k \gtrsim 1\,h\,$Mpc$^{-1}$, see Appendix~\ref{subsec:HMcode} and Figure~\ref{fig:pyhmcode_inaccuracy}), we use the CAMELS project to formulate the $P_\mathrm{ee}(k,z)$ in terms of cosmological and astrophysical feedback parameters that govern the cumulative feedback energy in hydrodynamical simulations. We calibrate $P_\mathrm{ee}(k,z)$ using six parameters in the $LH$ dataset variations of \texttt{IllustrisTNG} simulation suite. We show that the $P_\mathrm{ee}(k,z)$, suppression in the matter power spectrum, $P_\mathrm{hydro}/P_\mathrm{gravity-only}$ and the variance in FRB DM$_\mathrm{cosmic}$ are correlated with feedback parameters in hydrodynamical simulations (see Figure~\ref{fig:baryonic_feedback}). We also show that the variations of \texttt{IllustrisTNG} in the $LH$ dataset range from approximately no feedback scenario with zero suppression of the matter power spectrum to suppression to 25\% in the strong feedback scenario (see Figure~\ref{fig:baryonic_feedback_LH_TNG}). This wide range of variations makes it a useful dataset to calibrate the $P_\mathrm{ee}(k,z)$ and hence, the variance in FRB DM$_\mathrm{cosmic}$, as a function of feedback parameters.

    \item We validate our DM$_\mathrm{cosmic} - z$ analysis framework by performing MCMC analyses on a mock sample of $10^4$ FRBs drawn from the fiducial run of the \texttt{IllustrisTNG} simulation. We find that our model recovers the true parameters and we forecast a $2\%$ constraint on $H_0$, 16\% constraint on $\Omega_\mathrm{m}$, 8\% constraint on $\sigma_8$ and 2\% constraint on $p(\mathrm{DM}_\mathrm{host})$ distribution parameters. Furthermore, although the variance in DM$_\mathrm{cosmic}$ integrates out the scale dependence of the $P_\mathrm{ee}(k,z)$, we are able to recover the shape of the $P_\mathrm{hydro}/P_\mathrm{gravity-only}$ due to the correlations with constrained feedback parameters. We show that with $10^4$ FRBs, the 68\% confidence interval of samples drawn from the posterior constrain the $P_\mathrm{hydro}/P_\mathrm{gravity-only}$ within 0.6\%, 2\%, 8\% and 15\% at scales $k=0.5,\,1,\,10$ and $50\,h\,$Mpc$^{-1}$, respectively. With a realistic sample of $10^2$ FRBs (already discovered, localized and associated with their host galaxies!), the $P_\mathrm{hydro}/P_\mathrm{gravity-only}$ can be constrained within 1\%, 3\%, 12\% and 20\% at scales $k=0.5,\,1,\,10$ and $50\,h\,$Mpc$^{-1}$, respectively.

    \item We illustrate the flexibility of our formulation by demonstrating its ability to recover the feedback scenarios implemented in \texttt{SIMBA} and \texttt{Astrid} simulation suites, which are widely different from \texttt{IllustrisTNG} (see Figure~\ref{fig:CAMELS_simulations}, \ref{fig:baryonic_feedback_all_sims} and \ref{fig:DM_z_sims}). The 68\% confidence interval of samples drawn from the posterior constrain $P_\mathrm{hydro}/P_\mathrm{gravity-only}$ in \texttt{Astrid} within 0.5\%, 2\%, 18\% and 32\% at scales $k=0.5,\,1,\,10$ and $50\,h\,$Mpc$^{-1}$, respectively (see Figure~\ref{fig:suppression_varDM_results}). Similarly, in \texttt{SIMBA}, the recovery accuracy is 2.5\%, 8\%, 16\% and 22\% at scales $k=0.5,\,1,\,10$ and $50\,h\,$Mpc$^{-1}$, respectively (see Figure~\ref{fig:suppression_varDM_results}). The visible differences at large scales for \texttt{SIMBA} can be attributed to the finite box size of CAMELS project. The ability to measure the $P_\mathrm{hydro}/P_\mathrm{gravity-only}$ can allow us to infer the halo gas profiles as parameterized by the standard halo models.

    \item While this approach offers a promising path towards parameterizing $p(\mathrm{DM}_\mathrm{cosmic}|z)$, there are certain limitations that must be addressed before performing analyses with observed FRB data. The variance calibration needs to be updated to account for variance from large scales, which can be as large as 10\% of the variance when integrating down to $k \sim 0.01\,h\,\mathrm{Mpc}^{-1}$ (see Figure~\ref{fig:power_spectra_correlation_fxn_gas_cdm}). The strong correlation of $H_0$ and $\Omega_\mathrm{b}$ with other cosmological and feedback parameters suggest these should be incorporated in the calibration dataset when fitting for variance (see Figure~\ref{fig:correlation_matrix}). Furthermore, the current interpolation method used in a 6D parameter space with only 1000 data points is sparse, suggesting the need for dimensionality reduction or surrogate modeling techniques. Additionally, the framework developed for mock FRBs requires adjustments to account for interferometer detection efficiency in real-world observations~\citep{2022MNRAS.509.4775J, 2022MNRAS.516.4862J}. Lastly, while the log-normal approximation is used for the density field distribution, future work may explore likelihood-free inference to quantify deviations from log-normality~\citep{2024arXiv241007084K}.
    
\end{itemize}

We have demonstrated the sensitivity of FRBs to the matter power spectrum, and therefore to the effects of baryonic feedback, on scales up to $k \sim 10\,h\,$Mpc$^{-1}$. FRB constraints on the strength of baryonic feedback can significantly impact galaxy formation models. FRB constraints can inform weak lensing measurements of the next generation of observational cosmology experiments by mitigating baryonic uncertainties in cosmological analyses.

\begin{acknowledgments}
\nolinenumbers
We thank Matthew McQuinn for their comments on the manuscript. K.S. thanks Matthew McQuinn, Dhayaa Anbajagane, Volker Springel, Clancy James, Xavier Prochaska, Daohong Gao, Zijian Zhang and Tilman Tröster for insightful conversations that have helped shape this paper.

\end{acknowledgments}

\appendix

\section{Sensitivity of Baryon Probes}\label{appendix:halo_sensitivity}

To first order, the sensitivity can be computed as the product of the halo mass function $n(M)$ and the halo mass scaling relation of the respective probe. For example, the halo profile of X-rays emitted by bremsstrahlung scales as $\sim M^2 T^{1/2} \propto M^{7/3}$~\citep{2024arXiv241212081L, 2025arXiv250400113S}, while that of the tSZ effect arising from Compton scattering of CMB photons from hot gas in galaxy clusters scales as $\sim M T \propto M^{5/3}$. For reference, we also show the limits of eROSITA's first All-Sky Survey (eRASS1) galaxy clusters sample~\citep{2024A&A...689A.298G}, the South Pole Telescope (SPT) galaxy clusters sample~\citep{2024PhRvD.110h3510B}, and the kSZ measurements from the Atacama Cosmology Telescope (ACT) stacked on luminous red galaxies~\citep{2021PhRvD.103f3513S, 2024MNRAS.534..655B, 2024arXiv240707152H} in Figure~\ref{fig:sensitivity}.

For the key FRB observable, we quantify sensitivity as the fractional contribution of a differential halo mass bin centered at a given halo mass to the variance in DM. We use the halo model of \citet{2021MNRAS.502.1401M}, which assumes that ejected gas traces dark matter at large scales, as well as a more physical description of the ejected gas according to the gas profiles of \citet{2019JCAP...03..020S}, as implemented in \sw{BaryonForge}~\citep{2024OJAp....7E.108A}. The relative contribution of different mass scales depends on model parameters governing the fraction of bound gas and halo concentration (the shaded region in Figure~\ref{fig:sensitivity} denotes variations of these parameters). Similarly, we evaluate the halo mass sensitivity of weak lensing using the matter power spectrum. 

\begin{figure}
\centering
\includegraphics[width=\columnwidth]{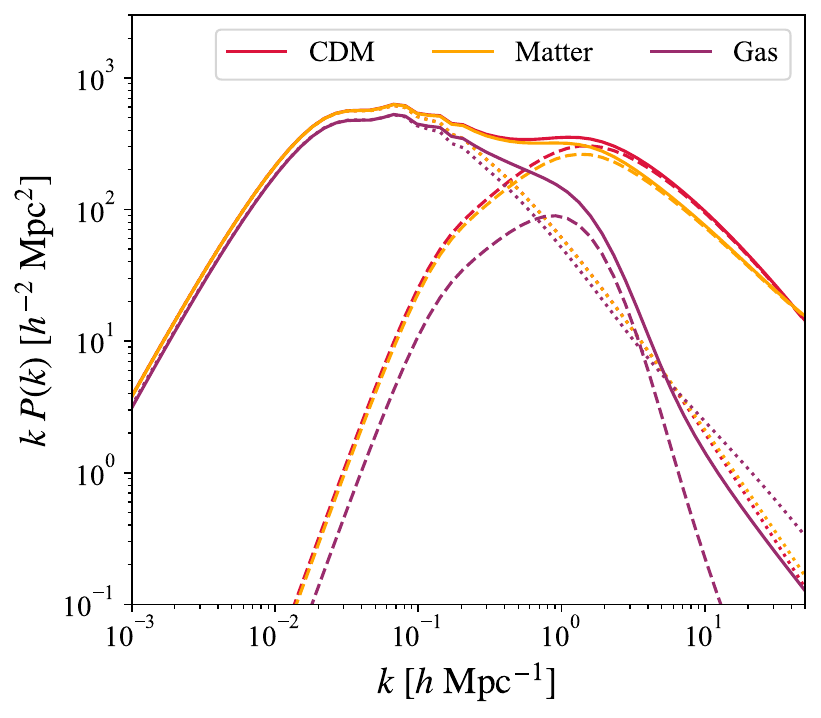}
\caption{The halo-model predictions for matter, cold dark matter (CDM), and gas power spectra~\citep{2020A&A...641A.130M, 2021MNRAS.502.1401M}. We rescale the halo model predictions (which normalizes the density using the mean matter density) to normalize with respect to the mean density in the respective field. Unlike the CDM field, where the two-halo term (dotted lines) is suppressed relative to the one-halo term (dashed lines) at small scales (depending on the halo profile and halo bias), the two-halo term for the gas component is not suppressed at small scales, leading to an nonphysically large contribution of the two-halo term compared to the one-halo term. This is because the expelled gas is approximated as a simple multiplicative bias at large scales, with no suppression at small scales. Secondly, the expelled gas is assumed to contribute only to the two-halo term, which is inaccurate, since the expelled gas from a halo must still be correlated with that halo, thus contributing to the one-halo term.}
\label{fig:pyhmcode_inaccuracy}
\end{figure}

\begin{figure*}
\centering
\includegraphics[width=\textwidth]{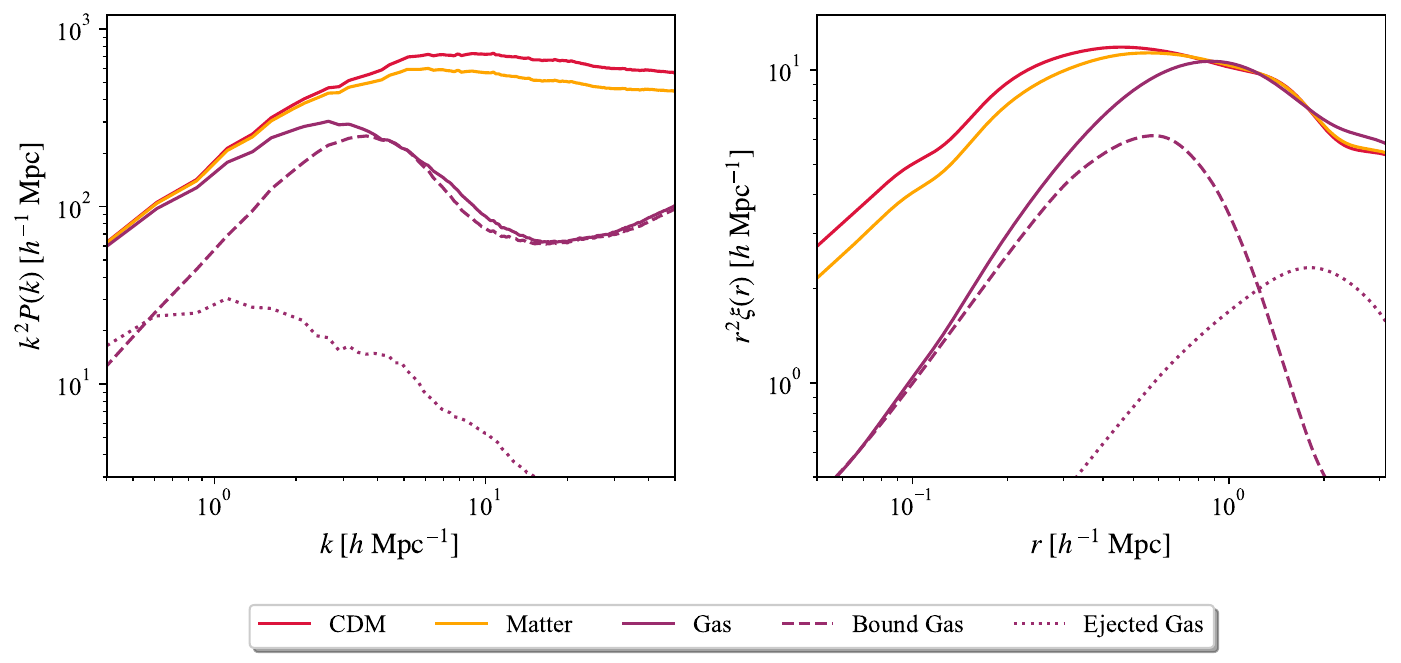}
\caption{The power spectra (left panel) in Fourier space and the autocorrelation function in configuration space for different fields, including matter, cold dark matter (CDM) and gas (similar to Figure~\ref{fig:power_spectra_correlation_fxn_gas_cdm}). In addition, we separately show the power spectra and the autocorrelation function of the bound gas (dashed lines) and the ejected gas (dotted lines). The majority contribution at large scales comes from the ejected gas component, which does not exactly trace dark matter.}
\label{fig:ejected_bound_gas}
\end{figure*}

\section{Halo Model at Small Scales} \label{subsec:HMcode}

Having identified the physical scales that contribute to the variance in DM$_\mathrm{cosmic}$, we evaluate the accuracy of the halo model predictions for the gas power spectrum at relevant scales for the variance in FRB DM~\citep[\texttt{HMcode};][]{2020A&A...641A.130M, 2021MNRAS.502.1401M}. The halo model framework involves three key ingredients: the radial density profile, the halo mass function, and the clustering bias of halos for each mass component. The expression for the auto-power spectrum of a mass field $u$ naturally decomposes into two terms, given by $P_\mathrm{uu}^{\mathrm{HM}}(k) = P_\mathrm{uu}^{\mathrm{2h}}(k) + P_\mathrm{uu}^{\mathrm{1h}}(k)$, where the two-halo term $P_\mathrm{uu}^{\mathrm{2h}}(k)$ accounts for the large-scale clustering of halos, while the one-halo term $P_\mathrm{uu}^{\mathrm{1h}}(k)$ represents the clustering of mass elements within individual halos at small scales. These terms are explicitly defined as follows:
\begin{equation}
\begin{aligned}
    P_\mathrm{uu}^{\mathrm{2h}}(k) &= P_\mathrm{uu}^{\mathrm{lin}}(k) \left[ \int\limits_0^\infty b(M) W_\mathrm{u}(M, k) n(M) \mathrm{d}M \right]^2, \\
    P_\mathrm{uu}^{\mathrm{1h}}(k) &= \int\limits_0^\infty W_\mathrm{u}^2(M, k) n(M) \mathrm{d}M,
\end{aligned}
\label{eqn:halo_model}
\end{equation}
where $P_\mathrm{uu}^{\mathrm{lin}}(k)$ denotes the linear theory prediction for the power spectrum, $M$ is the halo mass, and $b(M)$ is the linear halo bias, which relates the halo over-density $\delta_{\mathrm{h}}(M, k \to 0)$ to the linear matter over-density $\delta_{\mathrm{m}}(k \to 0)$ at large scales via $\delta_{\mathrm{h}}(M, k \to 0) = b(M) \delta_{\mathrm{m}}(k \to 0)$. Additionally, $n(M)$ represents the halo mass function, and $W_\mathrm{u}(M, k)$ is the spherical Fourier transform of the halo density profile.

To compute the power spectrum of the gas field, \citet{2020A&A...641A.130M, 2021MNRAS.502.1401M} decompose it into two distinct components: gas bound within halos (inside the virial radius) and gas ejected from halos (outside the virial radius). The bound gas component contributes to both the one-halo and two-halo terms, with its radial profile well described by the density profile of \citet{2001MNRAS.327.1353K}, which accurately characterizes halos unaffected by significant feedback processes. Conversely, the modeling of the ejected gas component is more complex. As a simplifying approximation, the ejected gas is assumed to contribute solely to the two-halo term, where it is treated as a biased tracer of the linear matter density field at large scales.

The halo model predictions for the power spectra of CDM, matter and gas fields are presented in Figure~\ref{fig:pyhmcode_inaccuracy}. At large scales, the spectra exhibit similar behavior, as the two-halo term ensures that the field distribution follows the large-scale linear perturbations in a biased manner. However, at small scales, the shapes differ significantly due to the distinct density profiles governing each component. While the one-halo term dominates at small scales for the CDM and matter power spectra, the gas power spectrum remains primarily governed by the two-halo term, even at small scales. 

This dominance of the two-halo term at small scales in the gas power spectrum may arise from two key factors. First, the one-halo term may be underestimated, as ejected gas remains in an extended shell beyond the halo's virial radius and retains correlations with the halo itself. Second, while the approximation of the two-halo term as a multiplicative bias is valid at large scales, it becomes inaccurate at small scales due to the expected suppression relative to linear theory predictions. Unlike the CDM field, where the two-halo term is appropriately suppressed according to the halo profile and halo bias, the gas field lacks this suppression. The combination of these effects leads to an overestimation of the gas power spectrum at small scales ($1 \lesssim k \lesssim 100\,h\,$Mpc$^{-1}$), a regime that is relevant for quantifying the variance in $\mathrm{DM}_\mathrm{cosmic}$.

We further emphasize the importance of accurately modeling both, the bound and the ejected gas components, using their power spectra and correlation functions, computed in the fiducial run of \texttt{IllustrisTNG}, as shown in Figure~\ref{fig:ejected_bound_gas}. The gas particles bound to halos are identified in the simulations using the halo catalogs constructed using the friends-of-friends algorithm. The total gas power spectrum is the sum of the bound gas auto-power spectrum, ejected gas auto-power spectrum and the bound gas-ejected gas cross-power spectrum (which is not shown here). We note that the bound gas dominates the contribution to the gas power spectrum at small scales. However, due to the small box size of the CAMELS project, these power spectra do not quite yet include the full two-halo term.

Application to upcoming datasets will require improved accuracy of halo model based predictions and there are several avenues for doing so. Models such as those presented in \citet{2015JCAP...12..049S}, \citet{2019JCAP...03..020S} and \citet{Arico2020} have more sophisticated prescriptions, especially for the gas component, and have been shown to match measurements from simulations. These models have already been applied to existing kSZ/tSZ/Xray datasets for constraining  suppression in matter power spectrum \citep[e.g.][]{2021JCAP...12..046G, 2022MNRAS.514.3802S, 2024MNRAS.528.4379G, 2025arXiv250116983B}, demonstrating their ability to physically link the baryonic fields to the total matter distribution. Both, large-volume cosmological simulations (e.g. \texttt{FLAMINGO}~\citep{2023MNRAS.526.4978S}, \texttt{BAHAMAS}~\citep{2017MNRAS.465.2936M, 2018MNRAS.476.2999M}, \texttt{Magneticum}~\citep{2014MNRAS.442.2304H, 2016MNRAS.456.2361B, 2025arXiv250401061D} and zoom-in simulations~\citep[e.g.][]{2018MNRAS.480..800H, 2020MNRAS.495.2930L}, can be used to update modeling ingredients and test their robustness. Recent work has sought to address some of the well known challenges with the halo model formalism, such as more accurate modeling of the non-linear halo bias which is especially important at the transition scale of the one- and two-halo terms ~\citep{2021MNRAS.502.1401M}.

\bibliography{manuscript}{}
\bibliographystyle{aasjournal}

\end{document}